\documentstyle[epsfig,graphics]{mn2e}

\title[A search for massive galaxies at ${\bf z > 4}$]{A systematic search 
for very massive galaxies at ${\bf z > 4}$}
\author[J.S. Dunlop et al..]
{J.S. Dunlop$^{1}$\thanks{jsd@roe.ac.uk}, 
M. Cirasuolo$^{1}$, R.J. McLure$^{1}$. 
\\
$^{1}$SUPA\thanks{Scottish Universities Physics Alliance}, 
Institute for Astronomy, University of Edinburgh, Royal Observatory,
Blackford Hill, Edinburgh, EH9 3HJ, UK.}

\date{Submitted for publication in MNRAS}

\begin{document}

\maketitle

\label{firstpage}

\begin{abstract}
Motivated by the claimed discovery of a 
very massive galaxy (HUDF-JD2; $M \simeq 5 \times 10^{11}\, 
{\rm M_{\odot}}$) at extreme redshift ($z = 6.5$) within the Hubble 
Ultra Deep Field (HUDF) (Mobasher et al. 2005), 
we have completed a systematic search for comparably 
massive galaxies with $z > 4$ among 
the 2688 galaxies in our $K_S < 23.5\, (AB)$ 
catalogue within the CDFS/GOODS-South 
field. This search was conducted using redshift estimates based on the 
recently-completed, uniquely-deep 11-band 
($B,V,i,z,J,H,K_S,3.6\mu m,4.5\mu m,5.8\mu m,8.0\mu  m$) 
imaging in this 125 square 
arcmin field, $\simeq 25$ times larger than the NICMOS HUDF. 
To ensure completeness, our approach places no special emphasis on the 
standard $V$-drop,
$i$-drop or $z$-drop criteria commonly used to pre-select 
candidate high-redshift
galaxies. 

Initial spectral fitting, based on published catalogue 
SExtractor photometry, led
us to conclude that at least 
2669 of the galaxies in our sample lie at $z < 4$.
This list includes several galaxies for which redshifts $z > 4$ 
have been previously proposed.
We carried out a 
detailed investigation of the 19 remaining $z > 4$ candidates, 
performing 
aperture photometry on all images, and including marginal detections 
and formal non-detections in the fitting process. This led to the rejection
of a further 13 galaxies to lower redshift.
Moreover, subjecting HUDF-JD2 to the same 
analysis, we find that it lies at $z \simeq 2.2$, 
rather than the extreme redshift favoured by Mobasher et al. (2005).

The 6 remaining candidates appear to be credible examples of galaxies 
in the redshift range $z = 4 - 6$, with plausible stellar ages. 
However, refitting with 
allowance for extreme values of
extinction we find that, even for these 
objects, statistically 
acceptable solutions can be found at $z < 3$. In fact only 
2 galaxies retain formally preferred high-redshift solutions.
Moreover, the recently-released Spitzer MIPS imaging in GOODS-South has 
revealed that 5 of our 6 final $z > 4$ candidates are detected at
24$\mu m$. This was also the case for HUDF-JD2 (Mobasher et al. 2005), 
and provides further circumstantial
evidence in favour of the moderate-redshift dusty solutions. We conclude that 
there is no convincing evidence for any galaxy with $M > 3 \times
10^{11}\, {\rm M_{\odot}}$ and $z > 4$ 
within the 125 square-arcmin GOODS-South field. We briefly 
discuss the implications of this null result, and revised expectations for 
the much larger (0.8 sq. degree), and deeper near-infrared UKIDSS Ultra Deep
Survey now underway with WFCAM on the UKIRT. 
\end{abstract}


\section{Introduction}

Several hundred convincing galaxy candidates have now been uncovered at 
$z \simeq 5-6.5$ ({\it e.g.} Bunker \& Stanway 2004; 
Taniguchi et al. 2005; Ouchi et al. 2005; 
Shioya et al. 2005; Yan et al. 2005; Bouwens et al. 2006), with a few
(somewhat less convincing) galaxy candidates even reported at $z > 7$
(Bouwens et al. 2004). The discovery of such objects has been used to set 
interesting new constraints on the cosmic history of star-formation density
({\it e.g.} Stark \& Ellis 2005, Bouwens \& Illingworth 2006), but has 
not, as yet, presented a serious challenge to current theories of galaxy 
formation. This is because the masses of essentially all of these 
objects are relatively modest ($M \simeq 10^{10} {\rm M_{\odot}}$), and 
the observed large numbers of such objects are consistent with the 
predictions of at least some current galaxy-formation models (e.g. Nagamine
et al. 2006). 

By contrast, the discovery of even a small number of very massive galaxies at 
these extreme redshifts can 
present a stern challenge  for both semi-analytic and hydrodynamic 
galaxy-formation models. Indeed, given the steep decline 
in the predicted number density of high-mass halos at $z > 4$
(e.g. Somerville 2004), the discovery 
of a significant number density of very massive objects
at such redshifts has the potential to provide an 
interesting test of the now well-established paradigm of  
hierarchical structure growth within $\Lambda CDM$.

For this reason, two recent studies have generated a lot of interest. First,
Eyles et al. (2005), in their detailed study of 3 spectroscopically confirmed
Lyman-break galaxies at $z \simeq 5.5 - 6$, reported that these objects already
contained a substantial, evolved mass of stars, apparently formed at redshifts
as high as $z \simeq 7.5 - 13.5$. 
Second, Mobasher et al. (2005) presented apparently
convincing evidence for the existence of an extremely massive galaxy 
(HUDF-JD2; $M \simeq 5 \times 10^{11}\, {\rm M_{\odot}}$)
lying within the NICMOS Hubble Ultra Deep Field (HUDF) 
at the extreme redshift of
$z = 6.5$. Although this is just one object, its discovery within such a very 
small-area survey such as the HUDF is undoubtedly surprising and has 
already generated considerable interest ({\it e.g.} Panagia et al. 2005).

Motivated by the discovery of HUDF-JD2, 
we decided to revisit our existing redshift determinations 
for $K_S$-band selected galaxies in GOODS-South 
(Caputi et al. 2004; 2005; 2006), and to conduct a systematic search for the 
existence of {\it any} galaxies in this sample at very high redshift 
($z > 4$). The key point here is that 
HUDF-JD2 is sufficiently massive that it is relatively bright in the 
near-infrared, with $K_S = 23.9$. Thus it should be possible to detect 
comparable objects (less than
a factor of 1.5 more massive at $z \simeq 6.5$) 
within our complete $K_S < 23.5$
sample of 2898 objects (galaxies and stars)
in the GOODS-South field, which covers a solid angle 
$\simeq 25$ 
times larger than that subtended by the NICMOS HUDF. It is also clear that,
in this
extreme mass domain, the discovery of
even one object within the entire GOODS-South field 
would be extremely important.

Given the apparent evidence for some moderately evolved stellar populations
in high-redshift galaxies, and because we wished to search for any galaxies
over a relatively wide redshift range $4 < z < 8$, 
we decided not to base our search on strict colour criteria. Such criteria
are frequently adopted in the selection of high-redshift galaxies, in order
to ensure minimal contamination from low-redshift interlopers (e.g. Bunker 
\& Stanway 2004). 
However,
especially at $z = 4-5$, such clean selection inevitably comes at the 
expense of completeness, and introduces an inevitable bias in favour 
of the youngest, and hence lowest mass:light ratio galaxies. Instead, 
we updated the dataset ultilised by Caputi et al. (2006) with the 
addition of the recently released ISAAC $H$-band and complete 
Spitzer 4-band IRAC data, 
and then derived new redshift estimates for all 2688 galaxies in the 
$K_S < 23.5$ sample by fitting a range of single and double component
spectrophotometric models to the full 11-band optical-infrared photometry.

The full results of this process are described in Cirasuolo et al. (2006) 
in which we have utilised this new dataset to explore the cosmological 
evolution of the galaxy mass function out to $z = 4$. In the current 
paper we use this work simply as a starting point for the careful 
study of the relatively small subsample of galaxies 
which were found to have even marginally convincing solutions at $z > 4$.

The paper is structured as follows. In Section 2 we briefly describe 
how we have utilised the updated public dataset within the GOODS-South field
to create a revised, complete sample of 2688 galaxies with $K_S < 23.5$ and 
full, aperture-matched 11-band HST/ACS+VLT/ISAAC+Spitzer/IRAC 
photometry. In Section 3 we 
describe the redshift-estimation technique, and demonstrate that it yields
robust solutions at the correct redshifts for known, 
spectroscopically-confirmed 
galaxies at $z \simeq 5-6$. Then, in Section 4, we describe the 
results of applying this technique to isolate a maximal sample of 32
potential $z > 4$ galaxies within our GOODS-South sample,
(using 2.8-arcsec diameter aperture photometry)
which was then refined down to a subsample
of 19 serious candidates using smaller aperture SExtractor 
(Bertin \& Arnout 1996) measurements. Section 5 then presents the 
results of a detailed investigation of the multi-frequency 
data for these remaining 19 candidates, with manual aperture photometry
leading to the rejection of a further 13 objects to lower redshift.
In this section we also show that, by subjecting HUDF-JD2 
to the same treatment,
we find that it does not lie at $z = 6.5$ (as concluded 
by Mobasher et al. 2005), but is in fact a dusty, evolved galaxy at 
$z \simeq 2.15$.
The final interrogation 
of our remaining six candidate $z > 4$ massive galaxies 
in presented in Section 6. In particular, we explore the robustness 
of the derived redshifts when the assumed optical extinction 
is allowed to float up to values as large as $A_V = 6$. Finally, in Section 7
we briefly 
discuss the implications of our results in the context of theoretical
models of structure formation.

All optical and infrared magnitudes are given in the $AB$ system
(Oke \& Gunn 1983).
Masses and ages have been calculated assuming a cosmological model 
with $\Omega_M = 0.3$,
$\Omega_{\Lambda} = 0.7$, and $H_0 = 70\,{\rm km s^{-1} Mpc^{-1}}$.
 
\begin{figure*}
\vspace*{15cm}
\includegraphics{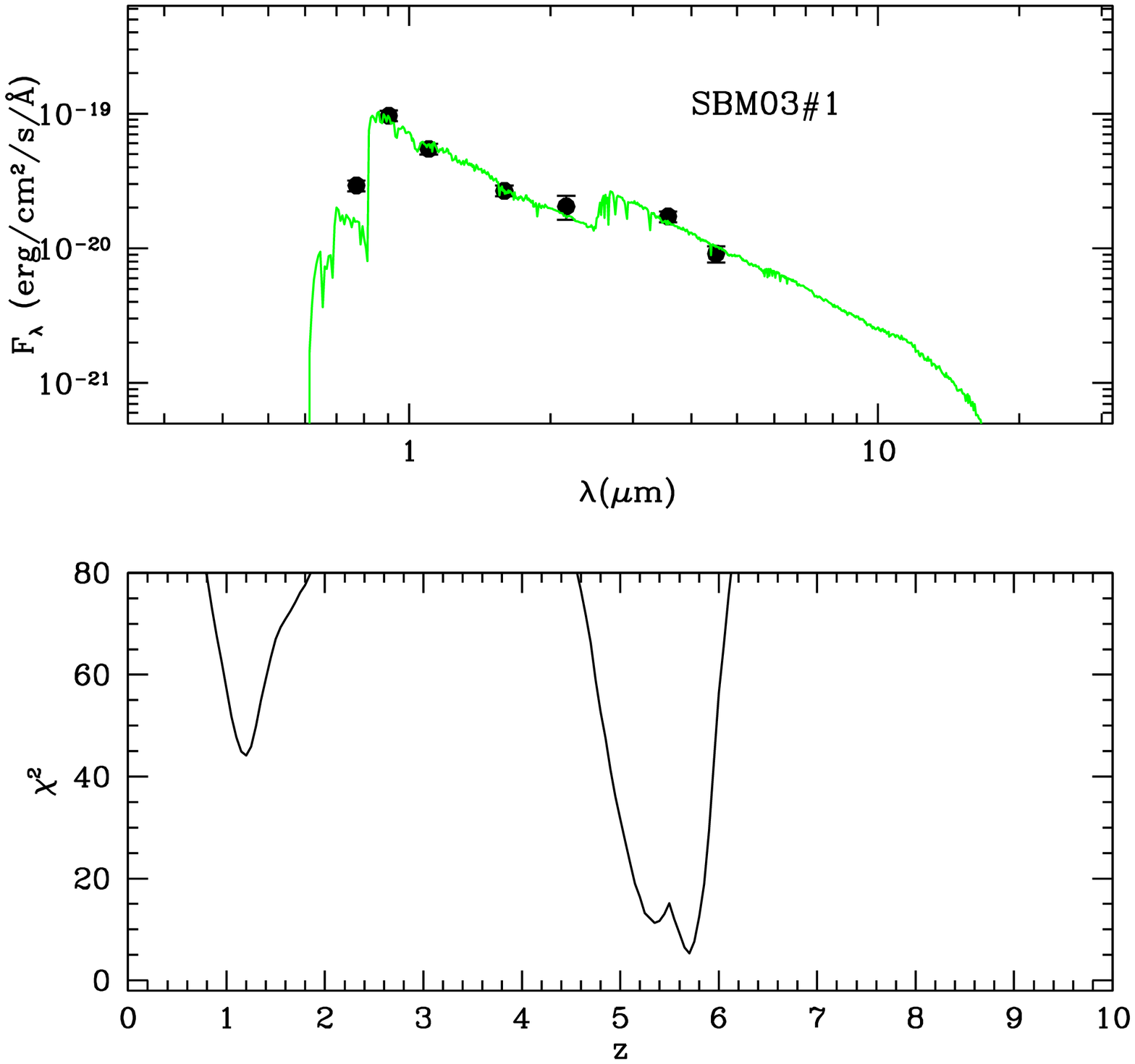}
\includegraphics{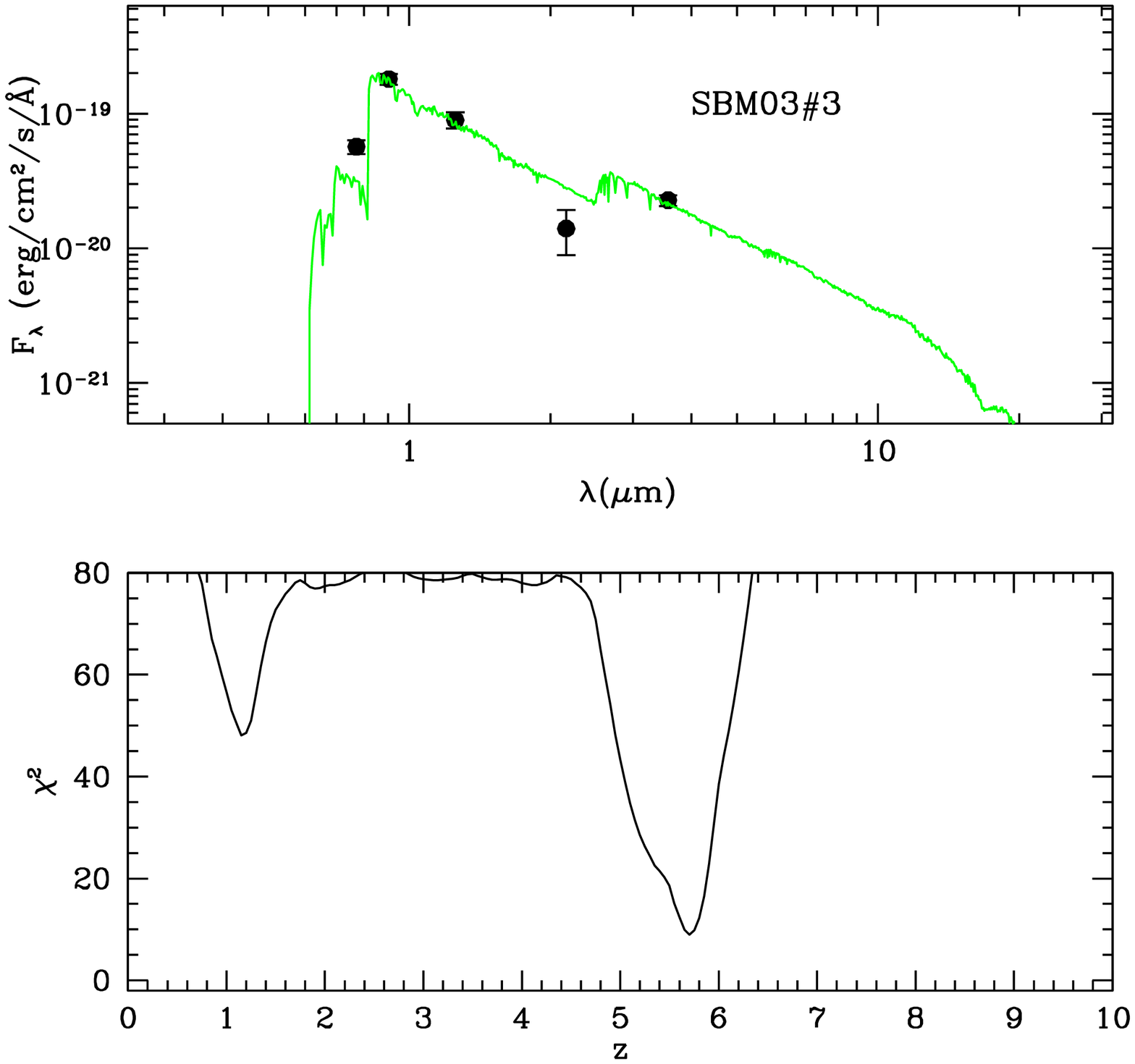}
\includegraphics{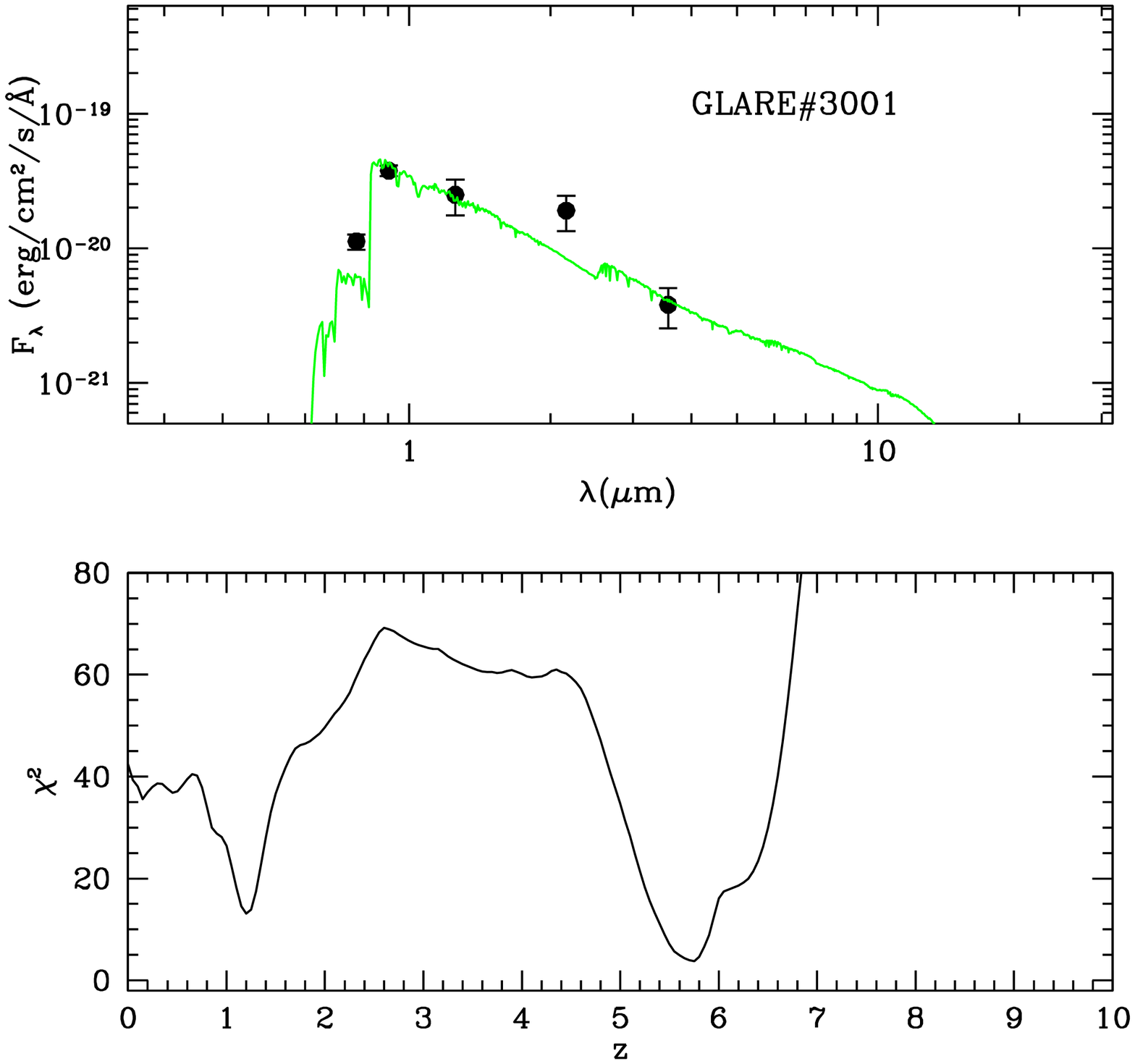}
\includegraphics{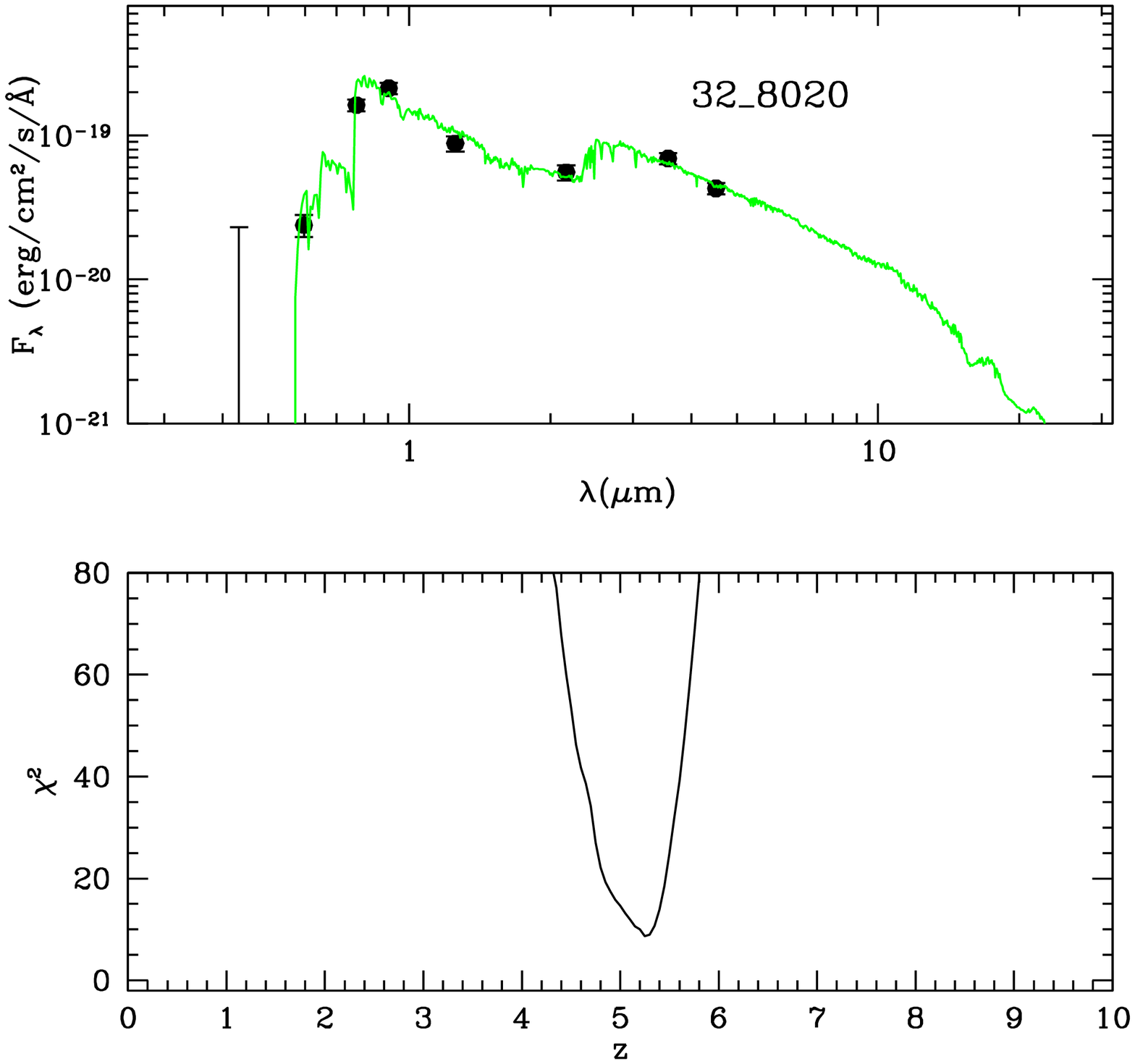}
\caption{\small Spectral fits and $\chi^2$ versus redshift $z$ for the
four galaxies in the HUDF and GOODS-South field with measured spectroscopic
redshifts $z > 5$, and published 
7-band photometry as tabulated by Eyles et al. (2005)
and Stark et al. (2006).
Details of the model fits, and a comparison of estimated and spectroscopic
redshifts are given in Table 1.}
\end{figure*}

\section{Data and sample definition}

\subsection{New public data}

The survey area available for this study is determined by the coverage of the 
VLT ISAAC $H$-band imaging of the GOODS-South field released in September 
2005. These data cover 125 arcmin$^2$, for which there already exists complete
ISAAC imaging in $J$, and $K_S$, along with complete HST ACS imaging in 
$B_{435}$ (3 orbits), $V_{606}$ (2.5 orbits), 
$i_{775}$ (2.5 orbits) and $z_{850}$ (5.0 orbits).

Spitzer IRAC imaging, in all 4 IRAC bands has now also become available 
for the whole of this field, with the 3.6\,$\mu m$ and 4.5\,$\mu m$ imaging 
considered by Caputi et al. (2006) now supplemented by the 5.8\,$\mu m$ and 
8.0\,$\mu m$ imaging. 

\subsection{Final sample properties}

The parent sample consists of the 2898 objects in this field which have 
$K_S \leq 23.5$ within a 2.8-arcsec diameter aperture. 
At this magnitude the limit
the sample should be essentially 100\% complete. 
Star-galaxy separation (SExtractor
stellaricity parameter $>0.8$ in the ACS $z$-band images) 
led to the rejection of 210 objects as stars,
leaving a final complete sample of 2688 galaxies.

Following several recent deep multi-object spectroscopy campaigns
(Vanzella et al. 2005; Doherty et al. 2005; Vanzella et al. 2006; Roche et al. 2006), 850 galaxies in our sample now possess
reliable spectroscopic redshifts. A further 188 galaxies,
with $R < 22.6$,  can be 
allocated solid redshifts from the COMBO-17 survey (Wolf et al. 2004). 
Thus, for 1650 galaxies we required to derive an 
estimated redshift based on the available, deep, 11-waveband photometry. 

\begin{table*}
\begin{center}
\caption{A test of a our redshift estimation code at high redshift.
Best fitting model parameters are given for our spectral fits 
to the broad-band photometry of the 
four spectroscopically confirmed $z = 5-6$ galaxies 
in the HUDF/GOODS-South field for which 7-band photometry 
has been published by Eyles et al. (2005) and Stark et al. (2006).
For each object we give the best fitting parameter values for both 
the best-fitting model with an exponentially declining star-formation rate, 
and 
for the best-fitting simple burst model. The fits are plotted in Figure 1, 
along with $\chi^2$ versus estimated redshift, 
marginalised over age, normalization, varied
star-formation history, and extinction.}
\begin{tabular}{lcccrccc}
\hline
Source & ${\rm z_{spec}}$ & ${\rm z_{est}}$ & Model type & $\chi^2$ &
Age / Gyr & ${\rm A_V}$ & Mass / $10^{10}\,{\rm M_{\odot}}$  \\
\hline
\\
SBM03\#1 & 5.83 & $5.70 \pm 0.07$ & $\tau = 0.3\,{\rm Gyr}$ & 5.26 
& 0.64 & 0.0 &\phantom{1}3.2\\
& & $5.40 \pm 0.15$ & Burst & 11.63 & 0.10 & 0.0 &\phantom{1}1.4\\
SBM03\#3 & 5.78 & $5.70 \pm 0.07$ & $\tau = 1.0\,{\rm Gyr}$ & 9.01 
& 0.91 & 0.0 & \phantom{1}6.5\\
&  & $5.55 \pm 0.15$ & Burst & 19.26     & 0.06 & 0.0 & \phantom{1}1.6\\
GLARE\#3001 & 5.79 & $5.75 \pm 0.12$ & $\tau = 0.3\,{\rm Gyr}$ & 3.80 
& 0.06 & 0.6 & \phantom{1}1.1\\
& & $5.62 \pm 0.15$ & Burst & 4.52 & 0.06 & 0.0 & \phantom{1}0.3\\
32\_8020 & 5.55 & $5.26 \pm 0.10$ & $\tau = 0.3\,{\rm Gyr}$ & 9.64
& 0.91 & 0.0 & 15.8\\
& & $4.76 \pm 0.15$ & Burst & 36.23 & 0.06 & 0.6 & \phantom{1}5.6\\
\hline
\end{tabular}
\end{center}
\end{table*}

\subsection{Initial multi-wavelength photometry}

The measurement of a robust 11-waveband spectral energy distribution for each 
source in the catalogue is non-trivial because the data span such 
a wide range in resolution (from $\simeq 0.1$ arcsec in the 
HST $B$-band images, 
through to $\simeq 2$  
arcsec in the Spitzer IRAC 8$\mu m$ imaging). To maximise sensitivity
at short wavelengths, it is 
tempting to consider the use of small ($< 1$ arcsec) apertures for 
source photometry in the ACS images. However, while most of the flux
from genuine high-redshift ($z > 4$) galaxies is likely to lie within a 
1 arcsec aperture, there are obvious dangers in combining small 
aperture HST photometry with the larger apertures required to obtain 
a robust estimate of total flux in the near-infrared ground-based imaging,
and the Spitzer IRAC imaging. In particular, unless sufficient 
care is taken with the aperture corrections, the strength of any putative 
break between the $J$ and $z_{850}$ filters can be exaggerated, 
leading to an erroneous conclusion in favour of a very high-redshift galaxy.

To minimize the effect of any such bias we based our initial photometric catalogue on the use of 2.8-arcsec diameter apertures for all the optical and 
near-infrared data. For the 4 Spitzer IRAC bands we also used a
2.8-arcsec diameter aperture but then applied an aperture 
correction to estimate the anticipated total flux for a point 
source (ranging from 0.55 mag. at 3.6$\mu m$ to 1.0 mag. at 
8$\mu m$). 
As we explain below, once we had used these data to isolate the subset 
of potential $z > 4$ galaxies, we considered alternative strategies to push 
the imaging closer to its photometric limit.

\section{Redshift estimation}

\subsection{Technique}

The photometric redshift for each galaxy was computed by 
fitting the 11 photometric data points (from the $B$-band to $8\mu m$) with
synthetic galaxy templates. These templates were produced
using the stellar population synthesis models of Bruzual
\& Charlot (2003), assuming a Salpeter
initial mass function (IMF) with a lower and upper mass cutoff at 0.1 and
100 $M_{\odot}$ respectively (we also explored the use of the Jimenez 
et al. (2004) models, and the Chabrier (2003) IMF, but these yielded 
inferior fits to the data). 

A range of templates was constructed based on different 
assumed star-formation histories. Specifically, we 
considered i) a single instantaneous starburst with passive evolution 
thereafter, ii) exponentially declining 
star-formation rates with $e$-folding
times in the range $0.3 \leq \tau (Gyr)\leq 15$, and iii) two-component burst
models (to cope with the possibility of more
stochastic star-formation histories). In all cases we adopted solar 
metallicity.

To account for the effects of dust reddening we adopted the obscuration law
of Calzetti et al. (2000). Initially we allowed $V$-band extinction to range
up to $A_V = 2$, but ultimately we explored the expanded range 
$0 \leq A_V \le 10$ (see section 6). We also added a prescription for 
the Lyman series absorption due to the HI clouds in the inter
galactic medium, following Madau (1995).

Some additional information was also utilised to exclude unreasonable redshift
solutions. First the photometric redshift was constrained to lie at 
$z \le 2$ if the source had been detected in the (relatively shallow) $U$-band 
imaging of the CDFS undertaken as part of the ESO Wide Field Survey 
(Arnouts et al. 2001). Second, a high-redshift solution was excluded 
if it resulted in a galaxy lying more than 1.3 magnitudes 
brightward of the $K-z$ relation defined by the most luminous radio galaxies 
(Willott et al. 2003), in effect equivalent to a galaxy more
massive than a present-day $10 L^{\star}$ elliptical.

\subsection{Tests} 

Our redshift estimation code has been tested for the $\simeq 1000$ galaxies 
in our GOODS-South $K_S$-selected sample which possess a reliable spectroscopic or 
COMBO-17 redshift. As detailed in Cirasuolo et al. (2006), the 
1$\sigma$ uncertainty in estimated redshift inferred from this test is 
$\delta z / (1 + z) \simeq 0.07$. 

However, of more specific interest for the present study is the ability 
of our method to accurately estimate the redshifts of known objects at 
$z > 4$. To check this, we applied our code to estimate the redshifts 
of the four spectroscopically confirmed $z \simeq 5 - 6$ galaxies
in the UDF/GOODS field. This is a useful test 
because the published multi-wavelength photometry for these 4 sources
(Eyles et al. 2005, Stark et al. 2006) 
provides comparable wavelength coverage  
to the photometry available for our complete GOODS-South sample.

The results of this test are summarized in Table 1, and 
illustrated in Figure 1 which shows, for each
galaxy, a plot of 
$\chi^2$ versus $z$ (marginalised over age, normalization, and $A_V$)
and a comparison of the best fitting model SED with the broad-band photometry.

Clearly, the code does an excellent job of recovering the redshifts of 
these galaxies. However, this is arguably a rather easy test, since these 
are, by 
nature of their selection, rather clear-cut examples of young Lyman-break
galaxies. As discussed further below, and as illustrated by the example of
HUDF-JD2 (Mobasher et al. 2005), the situation is inevitably more confused
if one is dealing with a potentially more evolved stellar population, with 
a less blue SED longward of the Lyman break. Nevertheless, the results shown in Figure 1  do at least confirm 
that our multi-wavelength SED fitting technique can 
efficiently and unambiguously identify high-redshift galaxies without recourse 
to any pre-selection of candidates based on, for example, 
specific colour criteria.

For completeness, we also note that the masses and ages 
we have derived for these galaxies are in excellent agreement with 
those derived by Eyles et al. (2005) and Stark et al. (2006).

\begin{table*}
\begin{center}
\caption{Estimated redshifts and other model parameter values for 
fits to the observed spectral energy distribution of HUDF-JD2 using 
i) the published 11-waveband photometry (Mobasher et al. 2005) and ii)
the revised photometry described in Section 5.2.}
\begin{tabular}{lccrccc}
\hline
Source & ${\rm z_{est}}$ & Model type & $\chi^2$ &
Age / Gyr & ${\rm A_V}$ & Mass / $10^{11}\,{\rm M_{\odot}}$  \\
\hline
\\
HUDF-JD2: published photometry & $6.20 \pm 0.20$ & Burst & 4.4
& 0.57 & 0.0 &6.0\\
HUDF-JD2: revised photometry & $2.15 \pm 0.30$ & Burst & 10.9
& 0.10 & 3.8 &0.8\\
\hline
\end{tabular}
\end{center}
\end{table*}

\begin{figure}
\vspace*{15cm}
\includegraphics{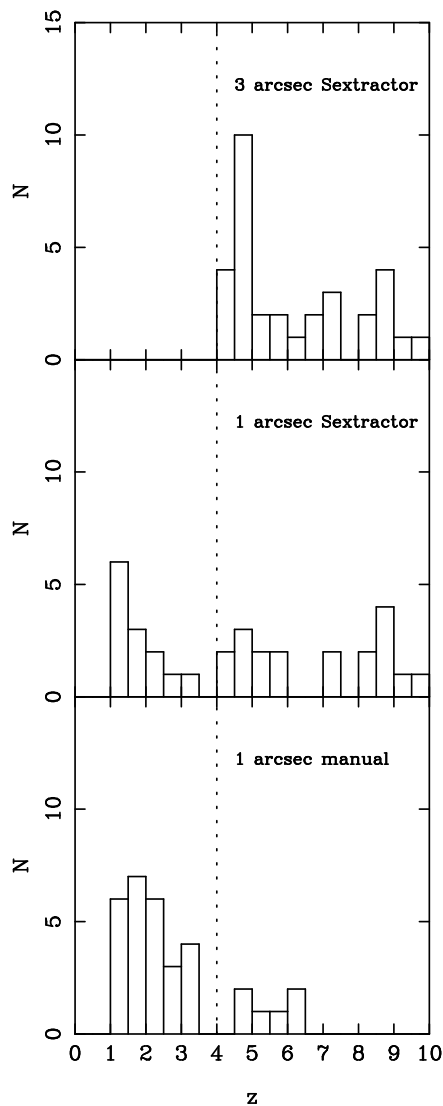}
\caption{\small Progressive refinement of the estimated 
redshift distribution of the initial 32-source $z > 4$ candidate list as 
the photometric measurements are pushed closer to the limit allowed by the 
imaging data.
The top panel shows the redshift distribution based on initial 2.8-arcsec 
diameter aperture magnitudes, and adopting formal limits for the non-detections
in the ACS optical bands. The middle panel shows the impact of 
moving to smaller (1-arcsec diameter) SExtractor 
magnitudes. The bottom panel shows the effect
of refitting the 19 objects which lie at $z > 4$ in the middle panel
to the manual aperture photometry tablulated in Tables 3 and 4.
The 6 objects which apparently remain at $z > 4$ are discussed in detail
in Section 5.3 and Section 6.}
\end{figure}

\section{Refinement of a high-redshift sample}

\subsection{Initial $z > 4$ sample}

Initial application of our redshift estimation code to the full 
2688-galaxy 11-waveband photometric dataset described above yielded 
formally acceptable solutions at $z > 4$ for 32 galaxies. This subset 
did not contain any of the objects for which redshifts have been 
obtained with the VLT (Vanzella et al. 2005, 2006), 
nor did it include any of the 18 claimed 
$V$-drop or $I$-drop galaxies within our sample listed by Bremer et al. (2004)
and Dickinson et al. (2003), all of which were found to lie at $z < 3$.
It also contains only 1 of the 7 galaxies within our sample which are 
listed as having $z > 4$ in the GOODS-MUSIC catalogue (Grazian et al. 2006).  

The initial derived redshift distribution for these 32 objects is shown in 
the top panel of Figure 2.
 
\subsection{Revised $z > 4$ sample after 1-arcsec catalogue search}

We next explored the 1-arcsec diameter aperture HST SExtractor catalogue
for these 32 sources. This was done to check if any non-detections 
in the larger aperture
became detections with reduced noise. The result of this was the relegation
of a further 13 galaxies to lower redshifts, leaving a reduced subset of 
19 potential $z > 4$ galaxies warranting further investigation. The result 
of this stage in the filtering is shown by the central panel in Figure 2, and 
details of the remaining 19 sources are given in Table 2.

\section{Detailed study of the final high-redshift sample}

\subsection{Direct aperture photometry}

At this point in the analysis, the values adopted for limiting magnitudes 
in the case of SExtractor non detections become crucial. In particular, the 
extent to which a high-redshift solution is favoured can be critically 
dependent on how any apparent non-detections are treated. 

One way to tackle this is the approach taken by Mobasher et al. (2005) 
in their analysis of HUDF-JD2. Faced with apparent non detection of this object
in the deep HUDF ACS optical imaging, they adopted 
$2\sigma$ upper limits of $B = 30.6$, $V = 31.0$, $i_{775} = 30.9$, $z_{850} 
= 30.3$, 
within their chosen 0.9-arcsec diameter aperture. We note here that 
at least some of these limits seem rather deep compared to
the limits adopted by some other authors working with 
HUDF data. For example Bouwens et al. 
(2004), in their search for $z_{850}$-drop galaxies in the HUDF, 
appear to have found a typical 2$\sigma$ limit of $z_{850} > 29.8$ within 
a 0.6-arcsec diameter 
aperture, which converts to $z_{850} 
> 29.4$ for a 0.9-arcsec diameter aperture,
almost a magnitude shallower than adopted by Mobasher et al. (2005).

We investigated the effect of calculating equivalent limits
for our GOODS non-detections ({\it i.e.} scaling the Mobasher et al. 
values to a 1-arcsec diameter aperture, and 
correcting for the GOODS:HUDF orbit ratio in each band). 
This leads to the conclusion
that our adopted $2\sigma$ limits should be 
$B = 28.9$, $V = 29.2$, $i_{775} = 28.6$, $z_{850} = 28.3$.

However, as illustrated in the central panel of Figure 2,
the adoption of such limits would lead
us to the conclusion that many of our remaining 19 sources lie at very high 
redshifts. Faced with such a radical conclusion we decided to abandon the 
inferred detection 
limits, and perform manual aperture photometry for all 19 galaxies 
to establish the true level of signal and noise, so that even the apparent
SExtractor non-detections could be properly incorporated within the 
$\chi^2$ fitting in a consistent manner. 

For consistency, we therefore decided to also 
subject HUDF-JD2 to the same type of analysis.

\begin{figure}
\vspace*{15cm}
\includegraphics{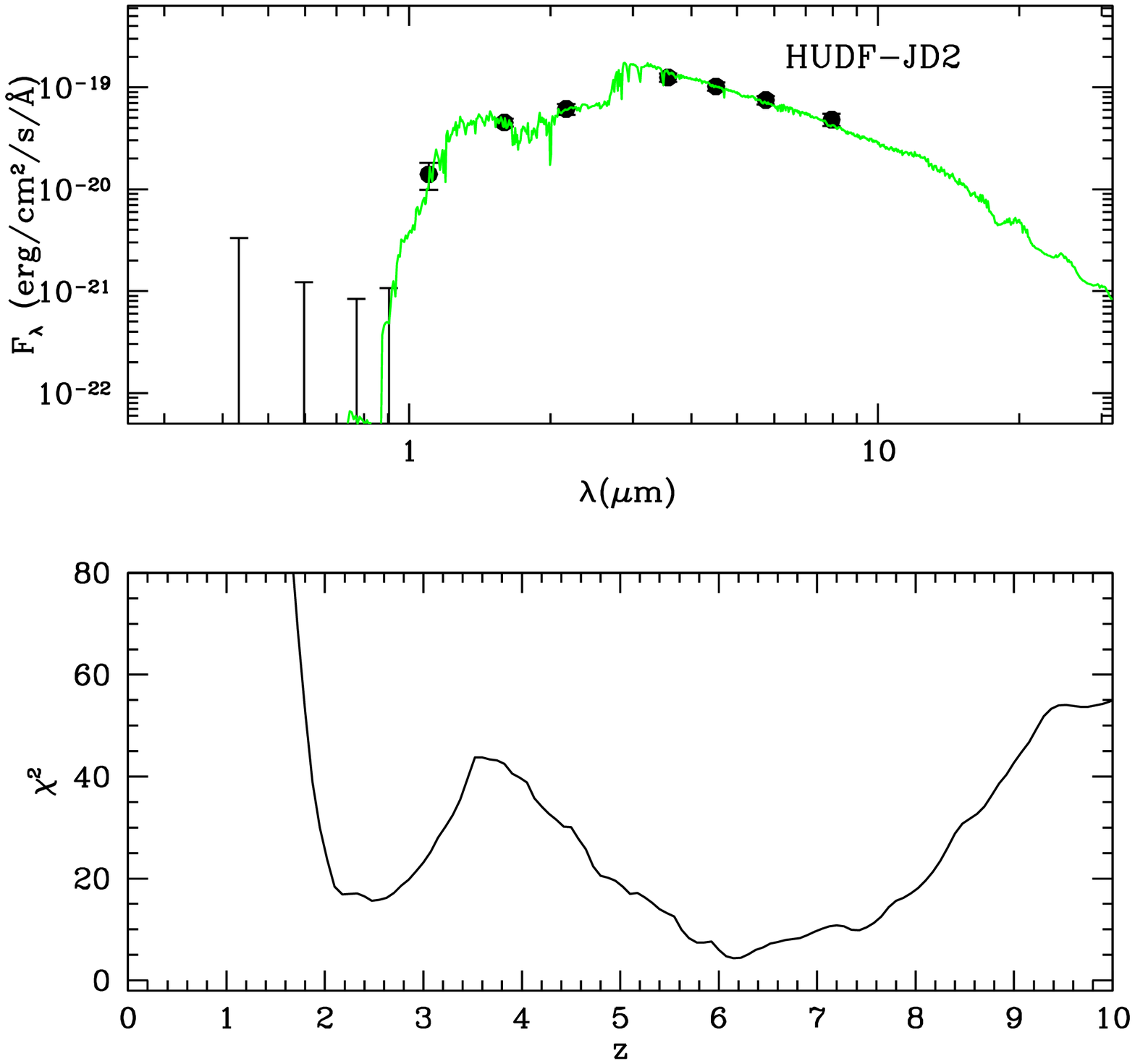}
\includegraphics{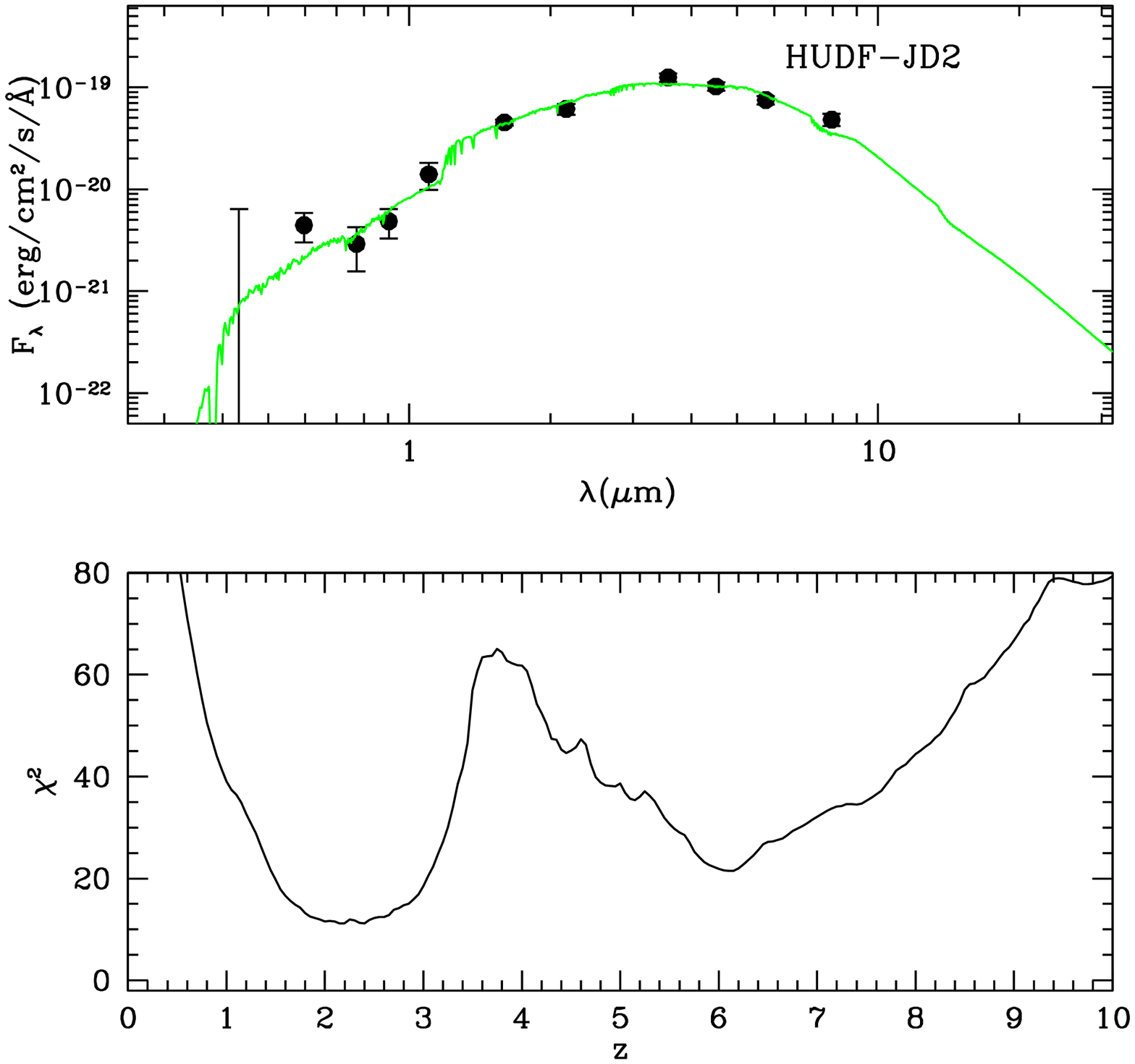}
\caption{\small Spectral fits and $\chi^2$ versus redshift $z$ for HUDF-JD2.
The top panel shows the result of applying our model fitting to the 
photometric data published by Mobasher et al. (2005). The lower panel 
shows the result of fitting to our own independent 
photometry, as discussed in Section 5.2. Best-fitting model parameters 
for both sets of data are given in Table 2.}
\end{figure}

\subsection{HUDF2 revisited}

In the top panel of Figure 3 we shown our own, independent fit to the 
photometry for HUDF-JD2 published by Mobasher et al. (2005). Our preferred
redshift of $z = 6.2 \pm 0.2$ is consistent with 
the value of $z = 6.5$ derived
by Mobasher et al. (2005), and we agree that the high-redshift solution 
is significantly favoured over the alternative option of a 
dust-obscured galaxy at $z \simeq 2 - 3$. The parameter values for our best
fit solution to the published photometry are given in Table 2.

However, performing our own 
manual photometry on the public HUDF images, through 
a 0.9-arcsec diameter aperture (as adopted by Mobasher et al.) we derive 
$B > 29.9$ (1$\sigma$), $V = 29.6 \pm 0.35$, $i_{775} = 29.5 \pm 0.5$ and 
$z_{850} = 28.6 \pm 0.35$. None of these highly 
marginal detections could, in isolation, be described as very convincing. 
However,  
their cumulative effect on the best fitting solution is dramatic, as 
illustrated in the lower panel of Figure 3. The best fitting result is now 
at $z = 2.15$, and the high-redshift 
solution is formally excluded. The parameter values for our best
fit solution to this revised photometry are also given in Table 2.
We conclude, therefore, that HUDF-JD2 lies at $z < 3$, 
and not at $z > 6$.

\begin{table*}
\begin{center}
\caption{Positions and 1-arcsec diameter aperture HST ACS 
optical magnitudes for the 
19 candidate $z > 4$ galaxies in our GOODS-South sample. Errors are given 
in magnitudes if the signal:noise ratio is greater than 3, and as a percentage 
error in flux density if the `detection' is less significant. 
1-$\sigma$ limits are given when the detected flux density was zero or negative
(see Section 5.3).}
\begin{tabular}{lccllll}
\hline
Name    & RA (J2000) & Dec (J2000) &\phantom{999}$B$  &\phantom{999}$V$   &\phantom{999}$i_{775}$    &\phantom{999}$z_{850}$\\
\hline
1865     &03 32 12.87 &$-$27 46 40.9 &\phantom{$>$}$29.3\ (180\%)$     &\phantom{$>$}$29.7\ (130\%)$
         &\phantom{$>$}$27.6\ (70\%)$      &\phantom{$>$}$26.8 \pm 0.30$\\
2028     &03 32 22.53 &$-$27 49 32.6&\phantom{$>$}$29.0\ (150\%)$     &\phantom{$>$}$27.0 \pm 0.20$
         &\phantom{$>$}$26.3 \pm 0.20$    &\phantom{$>$}$26.3 \pm 0.10$\\
2336     &03 32 25.25&$-$27 52 30.3&$>29.1$                         &\phantom{$>$}$28.8\ (80\%)$
         &\phantom{$>$}$27.9\ (55\%)$      &\phantom{$>$}$28.1\ (80\%)$\\
2351     &03 32 54.75&$-$27 51 13.8&$>28.7$                         &\phantom{$>$}$29.0\ (160\%)$
         &\phantom{$>$}$28.4\ (160\%)$     &\phantom{$>$}$27.5\ (100\%)$\\
2476     &03 32 37.86&$-$27 52 01.3&$>28.8$                         &$>29.0$
         &\phantom{$>$}$26.8 \pm 0.25$    &\phantom{$>$}$26.1 \pm 0.20$\\
2507     &03 32 19.67&$-$27 46 02.0&\phantom{$>$}$28.9\ (120\%)$     &$>29.5$
         &$>28.6$                         &\phantom{$>$}$27.8\ (65\%)$\\
2600     &03 32 38.34&$-$27 51 01.0&\phantom{$>$}$28.0\ (70\%)$      &\phantom{$>$}$27.8 \pm 0.35$
         &\phantom{$>$}$27.2 \pm 0.30 $   &\phantom{$>$}$27.0\ (45\%)$\\
2609     &03 32 56.10&$-$27 52 05.0&\phantom{$>$}$28.3\ (60\%)$      &\phantom{$>$}$28.1\ (50\%)$
         &\phantom{$>$}$27.1 \pm 0.18$    &\phantom{$>$}$26.3 \pm 0.20$\\
2694     &03 32 21.99&$-$27 51 11.9&\phantom{$>$}$28.3\ (70\%)$      &\phantom{$>$}$28.7\ (70\%)$
         &\phantom{$>$}$27.7\ (60\%)$      &\phantom{$>$}$28.9\ (200\%)$\\
2869     &03 32 17.99&$-$27 50 52.7&\phantom{$>$}$28.1\ (70\%)$      &\phantom{$>$}$27.25 \pm 0.25$
         &\phantom{$>$}$26.1 \pm 0.15$    &\phantom{$>$}$26.1 \pm 0.30$\\
2895     &03 32.16.83&$-$27 49 07.7&\phantom{$>$}$29.7\ (250\%)$     &\phantom{$>$}$28.6\ (50\%)$
         &\phantom{$>$}$27.6\ (60\%)$      &\phantom{$>$}$27.2\ (70\%)$\\
2957     &03 32 37.06&$-$27 44 19.1&\phantom{$>$}$28.8\ (180\%)$     &\phantom{$>$}$28.9\ (70\%)$
         &\phantom{$>$}$28.4\ (130\%)$     &\phantom{$>$}$26.9\ (55\%)$\\
2958     &03 32 42.08&$-$27 41 41.3&$>29.1$        &$>28.8$        &$>28.3$        &$>28.6$\\

3021     &03 32 27.14&$-$27 53 11.6&$>28.7$                         &\phantom{$>$}$28.2\ (75\%)$
         &\phantom{$>$}$26.7 \pm 0.35$    &\phantom{$>$}$26.8\ (35\%)$\\
3037     &03 32 28.21&$-$27 51 16.2&$>29.1$                         &\phantom{$>$}$27.3 \pm 0.20$
         &\phantom{$>$}$26.6 \pm 0.15$    &\phantom{$>$}$26.4 \pm 0.25$\\
3048     &03 32 19.57&$-$27 41 39.9&$>28.2$                         &$>29.1$
         &\phantom{$>$}$27.0\ (50\%)$      &\phantom{$>$}$26.8 \pm 0.33$\\
3087     &03 32 54.81&$-$27 51 38.8&\phantom{$>$}$27.8\ (40\%)$      &\phantom{$>$}$27.8\ (40\%)$
         &\phantom{$>$}$28.5\ (50\%)$      &\phantom{$>$}$26.8 \pm 0.25$\\
3088     &03 32 39.13&$-$27 51 05.0&\phantom{$>$}$28.5\ (120\%)$     &\phantom{$>$}$28.9\ (65\%)$
         &\phantom{$>$}$26.8 \pm 0.15$    &\phantom{$>$}$27.9\ (60\%)$\\
3122     &03 32 21.69&$-$27 42 42.3&$>29.1$                         &$\phantom{>}27.9\ (40\%)$
         &$\phantom{>}28.7\ (150\%)$     &$\phantom{>}27.9\ (80\%)$\\
\hline
\end{tabular}
\end{center}
\end{table*}

\begin{table*}
\begin{center}
\caption{Infrared photometry for the 19 candidate $z > 4$ galaxies. 
$J,H,K_S$ magnitudes have been rederived from the public VLT ISAAC imaging
using 1-arcsec diameter apertures, and applying a point-source
correction of $-0.5$ magnitudes. The Spitzer IRAC magnitudes 
have also been corrected to account for the point-spread function, and 
typical errors adopted for all objects at each waveband.}
\begin{tabular}{llllcccc}
\hline
Name     &\phantom{999}$J$   &\phantom{999}$H$     &\phantom{999}$K_S$ &$3.6\mu m$\phantom{22}&$4.5\mu m$\phantom{22}&$5.8\mu m$\phantom{22}&$8.0\mu m$\phantom{22}\\
\hline
1865     &\phantom{$>$}$24.33 \pm 0.13$   &\phantom{$>$}$23.52 \pm 0.06$
         &\phantom{$>$}$22.62 \pm 0.05$   &$20.96 \pm 0.20$&$20.72 \pm 0.20$ 
         &$20.64 \pm 0.30$ &  $21.02 \pm 0.30$\\
2028     &\phantom{$>$}$23.86 \pm 0.07$   &\phantom{$>$}$23.15 \pm 0.07$
         &\phantom{$>$}$22.66 \pm 0.04$   &$20.55 \pm 0.20$&$20.77 \pm 0.20$
         &$20.47 \pm 0.30$ & $20.60 \pm 0.30$\\
2336     &\phantom{$>$}$24.98 \pm 0.17$   &\phantom{$>$}$23.86 \pm 0.12$
         &\phantom{$>$}$22.96 \pm 0.04$   &$21.29 \pm 0.20$&$20.81 \pm 0.20$
         &$20.34 \pm 0.30$ & $20.30 \pm 0.30$\\
2351     &\phantom{$>$}$25.10 \pm 0.34$   &\phantom{$>$}$23.64 \pm 0.13$
         &\phantom{$>$}$22.98 \pm 0.07$   &$21.33 \pm 0.20$&$21.10 \pm 0.20$
         &$21.04 \pm 0.30$ & $21.26 \pm 0.30$\\
2476     &\phantom{$>$}$23.99 \pm 0.08$   &\phantom{$>$}$23.30 \pm 0.08$
         &\phantom{$>$}$22.96 \pm 0.05$   &$22.14 \pm 0.20$&$21.92 \pm 0.20$ 
         &$21.96 \pm 0.30$ & $22.26 \pm 0.30$\\
2507     &\phantom{$>$}$25.48 \pm 0.26$   &\phantom{$>$}$24.41 \pm 0.16$
         &\phantom{$>$}$23.04 \pm 0.04$   &$21.20 \pm 0.20$&$20.77 \pm 0.20$  
         &$20.25 \pm 0.30$ & $20.43 \pm 0.30$\\
2600     &\phantom{$>$}$25.20 \pm 0.22$   &\phantom{$>$}$23.83 \pm 0.09$
         &\phantom{$>$}$22.93 \pm 0.05$   &$21.59 \pm 0.20$&$21.26 \pm 0.20$ 
         &$21.23 \pm 0.30$ & $21.48 \pm 0.30$\\
2609     &\phantom{$>$}$24.63 \pm 0.14$   &\phantom{$>$}$23.42 \pm 0.08$
         &\phantom{$>$}$23.12 \pm 0.06$   &$21.78 \pm 0.20$&$21.47 \pm 0.20$  
         &$21.64 \pm 0.30$ & $21.71 \pm 0.30$\\
2694     &\phantom{$>$}$25.87 \pm 0.41$   &\phantom{$>$}$24.59 \pm 0.22$
         &\phantom{$>$}$23.08 \pm 0.05$   &$21.31 \pm 0.20$&$20.88 \pm 0.20$
         &$20.47 \pm 0.30$ & $20.78 \pm 0.30$\\
2869     &\phantom{$>$}$24.89 \pm 0.20$   &\phantom{$>$}$24.06 \pm 0.15$
         &\phantom{$>$}$23.18 \pm 0.06$   &$21.57 \pm 0.20$&$21.37 \pm 0.20$
         &$21.33 \pm 0.30$ & $20.41 \pm 0.30$\\
2895     &\phantom{$>$}$25.01 \pm 0.18$   &\phantom{$>$}$24.13 \pm 0.15$
         &\phantom{$>$}$22.99 \pm 0.04$   &$21.68 \pm 0.20$&$21.57 \pm 0.20$
         &$21.73 \pm 0.30$ & $21.85 \pm 0.30$\\
2957     &\phantom{$>$}$25.80 \pm 0.33$   &\phantom{$>$}$24.65 \pm 0.15$
         &\phantom{$>$}$23.05 \pm 0.04$   &$21.89 \pm 0.20$&$21.52 \pm 0.20$ 
         &$21.06 \pm 0.30$ & $21.32 \pm 0.30$\\
2958     &\phantom{$>$}$25.51 \pm 0.33$   &\phantom{$>$}$24.56 \pm 0.23$
         &\phantom{$>$}$23.34 \pm 0.09$   &$21.34 \pm 0.20$&$20.95 \pm 0.20$ 
         &$20.48 \pm 0.30$ & $20.44 \pm 0.30$\\
3021     &\phantom{$>$}$25.27 \pm 0.19$   &\phantom{$>$}$24.29 \pm 0.15$
         &\phantom{$>$}$23.46 \pm 0.07$   &$21.77 \pm 0.20$&$21.49 \pm 0.20$ 
         &$21.09 \pm 0.30$ & $21.19 \pm 0.30$\\
3037     &\phantom{$>$}$24.60 \pm 0.12$   &\phantom{$>$}$24.01 \pm 0.12$
         &\phantom{$>$}$23.39 \pm 0.07$   &$22.38 \pm 0.20$&$22.13 \pm 0.20$
         &$21.53 \pm 0.30$ & $22.18 \pm 0.30$\\
3048     &\phantom{$>$}$25.60 \pm 0.31$   &\phantom{$>$}$24.26 \pm 0.23$
         &\phantom{$>$}$23.49 \pm 0.08$   &$21.69 \pm 0.20$&$21.29 \pm 0.20$  
         &$20.77 \pm 0.30$ & $20.93 \pm 0.30$\\
3087     &\phantom{$>$}$25.30 \pm 0.29$   &\phantom{$>$}$24.16 \pm 0.17$
         &\phantom{$>$}$23.22 \pm 0.07$   &$21.83 \pm 0.20$&$21.55 \pm 0.20$
         &$21.01 \pm 0.30$ & $21.26 \pm 0.30$\\
3088     &\phantom{$>$}$24.60 \pm 0.12$   &\phantom{$>$}$24.19 \pm 0.15$
         &\phantom{$>$}$23.33 \pm 0.08$   &$21.92 \pm 0.20$&$21.64 \pm 0.20$  
         &$21.62 \pm 0.30$ & $22.03 \pm 0.30$\\
3122     &$\phantom{>}27.21 \pm 0.16$    &$\phantom{>}23.85 \pm 0.14$
         &$\phantom{>}23.28 \pm 0.09$     &$21.81 \pm 0.20$&$21.54 \pm 0.20$  
         &$21.20 \pm 0.30$ & $21.35 \pm 0.30$\\
\hline
\end{tabular}
\end{center}
\end{table*}

\subsection{Final photometry and redshift distribution}

The positions, and final 11-waveband photometry for the 19 
remaining $z > 4$ candidates are 
given in Tables 3 and 4. Optical magnitudes are based on manual photometry
through a 1-arcsec diameter aperture as described above, with errors given 
in magnitudes if the signal:noise ratio is greater than 3, and as a 
fractional error in flux if the `detection' is less significant. To avoid 
bias, the $J,H,K_S$ values have been re-derived also using a 1-arcsec 
diameter aperture,
with a minimal (effectively stellar) aperture correction of $-0.5$ magnitudes
applied to correct for missing flux.

The estimated redshifts resulting from fitting to the final 
photometry for these 19 galaxies are listed in Table 5, with the resulting 
redshift distribution shown in the bottom panel of Figure 2. Clearly all 19 
of these galaxies are interesting objects, with a sharp decline in flux-density
shortward of the $J$-band, but now only 6 galaxies retain credible solutions
at $z > 4$. Plots of $\chi^2$ versus redshift (marginalised over 
age, extinction and star-formation history) 
and the best-fitting model SEDs for these 6 galaxies
are shown in Figure 4.

For comparison, we have also listed in Table 5 the estimated redshifts for
these 19 galaxies recently released by Grazian et al. (2006) as part of the
GOODS-MUSIC project. We have also applied our own spectral fitting technique 
to the GOODS-MUSIC photometry for these objects, to allow us to explore the 
extent to which any disagreements may depend on photometry or model 
fitting. Inspection of Table 5 shows that, for many of these galaxies, the 
agreement between all three redshift estimates is excellent. However, there 
are also obvious differences. Specifically, only one of these
objects has a redshift $z > 4$ in the GOODS-MUSIC list, and our own solution
for this object with either photometry set lies at $z < 4$. 
Before discussing further the likely explanation for this apparent
disagreement at high redshift, we describe below the result of our final 
analysis of our own remaining subset of 6 potential $z > 4$ candidates.

\begin{table*}
\begin{center}
\caption{Re-estimated redshifts for the 19 candidate $z > 4$ galaxies in 
our GOODS-South sample. Column 2 gives the estimated redshift (and associated
uncertainty) which we have derived by applying the technique described in Section 3.1, to the revised photometric data tablulated in Tables 3 and 4. The values
of minimum $\chi^2$ for these fits are given in column 3. In columns 4 and 5 
we give the corresponding object IDs and estimated redshifts recently published
by Grazian et al. (2006), as part of the GOODS-MUSIC project. In column 6 
we give a third estimate of the redshift for each object, this time applying
our own technique to the photometric data published by Grazian et al. (2006). 
Column 7 gives the values of minimum $\chi^2$ for these fits, many of which 
are very large (apparently due to problems in fitting some of the claimed 
$U$-band detections in the GOODS-MUSIC catalogue).}
\begin{tabular}{llrrclr}
\hline
Source & ${\rm z_{est1}\,\,(1\sigma\, range)}$ & $\chi^2$ & MUSIC ID 
&MUSIC ${\rm z_{est}}$& ${\rm z_{est2}\,\,(1\sigma\, range)}$ & $\chi^2$\\
\hline
\\
1865  &5.02 (4.87-5.17)&  8.82 &  30093  &   2.04 & 2.00 (1.95-2.10)&72.03\\   
2028  &1.95 (1.80-2.10)& 42.30 &  30120  &   1.97 & 1.85 (1.80-1.95)&60.69\\
2336  &6.22 (6.07-6.45)& 12.73 &  30199  &   2.73 & 3.65 (3.25-3.85)& 2.77\\
2351  &1.88 (1.80-2.25)&  5.15 &  30142  &   1.84 & 1.85 (1.80-1.95)&25.34\\
2476  &1.65 (1.50-1.80)&  6.02 &  30160  &   1.59 & 1.70 (1.50-1.80)& 9.53\\
2507  &4.87 (4.72-5.10)&  2.20 &  30049  &   2.21 & 1.90 (1.70-2.20)& 2.13\\
2600  &2.18 (1.95-2.33)&  1.39 &  30106  &   2.05 & 2.15 (2.05-2.25)&78.35\\
2609  &2.33 (2.18-2.40)& 10.79 &  30146  &   2.06 & 2.15 (2.05-2.25)&27.48\\ 
2694  &2.93 (2.55-3.08)&  5.88 &  30114  &   2.57 & 3.45 (3.00-3.90)& 6.60\\
2869  &3.30 (3.08-3.45)& 10.08 &  30115  &   1.88 & 1.55 (1.45-1.65)&65.62\\
2895  &1.73 (1.58-1.88)&  4.42 &  30123  &   1.94 & 2.00 (1.75-2.45)& 4.24\\
2957  &3.45 (3.30-3.83)&  3.81 &  30048  &   4.66 & 3.75 (3.50-4.30)& 1.49\\ 
2958  &6.07 (4.95-6.30)&  1.49 &  30009  &   1.73 & 1.95 (1.35-2.35)& 0.71\\
3021  &3.00 (2.78-3.23)&  8.33 &  30175  &   2.44 & 2.20 (2.10-2.30)&21.98\\
3037  &2.25 (2.10-2.40)&  9.39 &   5177  &   2.73 & 1.75 (1.65-1.89)& 45.31\\  
3048  &4.87 (4.72-5.10)&  5.83 &  30036  &   2.62 & 2.45 (2.10-2.70)& 3.64\\ 
3087  &2.18 (1.88-2.33)& 12.44 &  30145  &   2.53 & 1.95 (1.75-2.10)&17.55\\ 
3088  &6.00 (5.85-6.37)& 14.14 &  30097  &   1.91 & 1.45 (1.40-1.55)&14.48\\
3122  &2.63 (2.25-2.85)&  8.98 &  30032  &   3.54 & 2.65 (2.45-2.85)&25.74\\
\hline
\end{tabular}
\end{center}
\end{table*}

\begin{figure*}
\vspace*{22.5cm}
\includegraphics{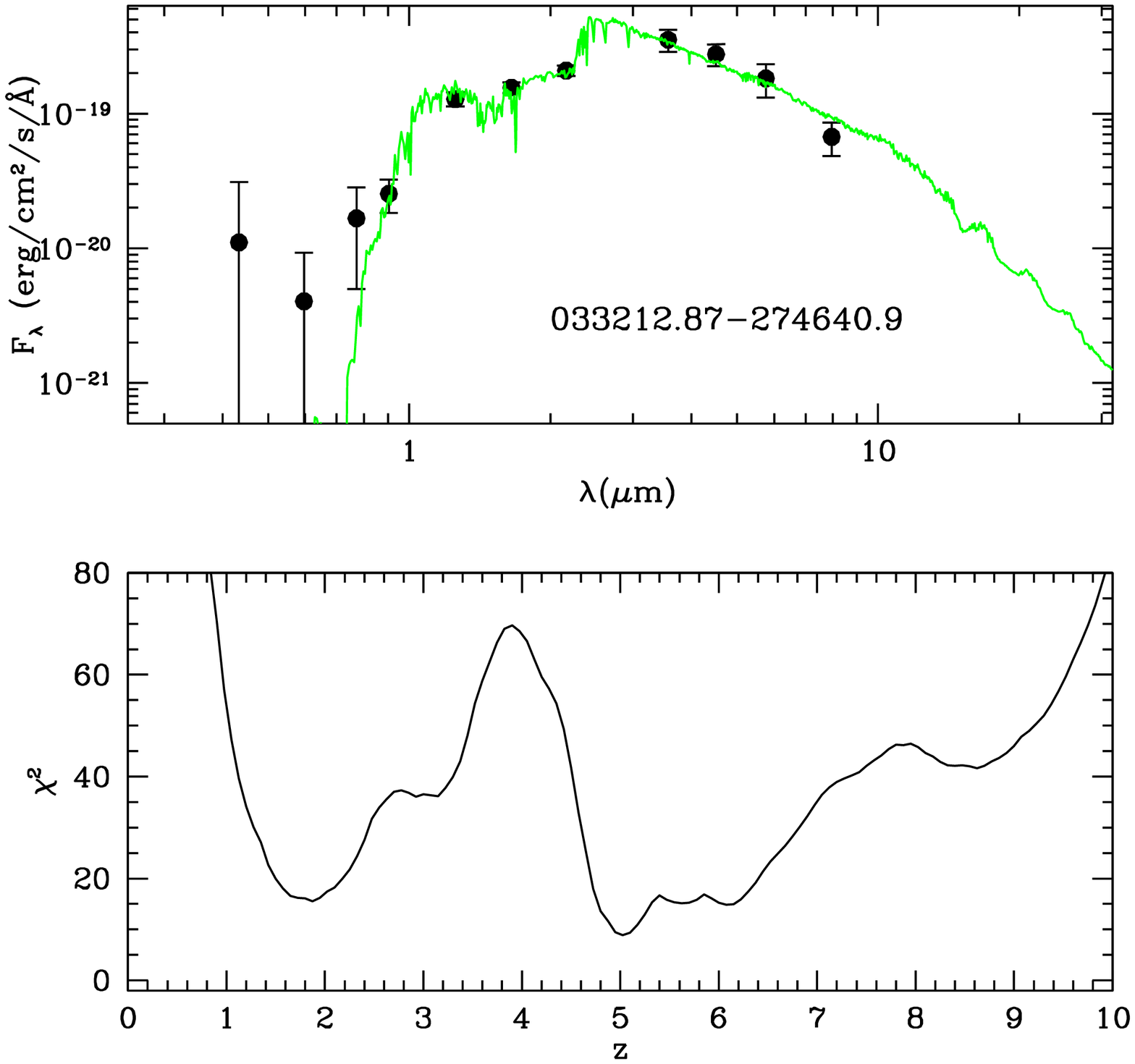}
\includegraphics{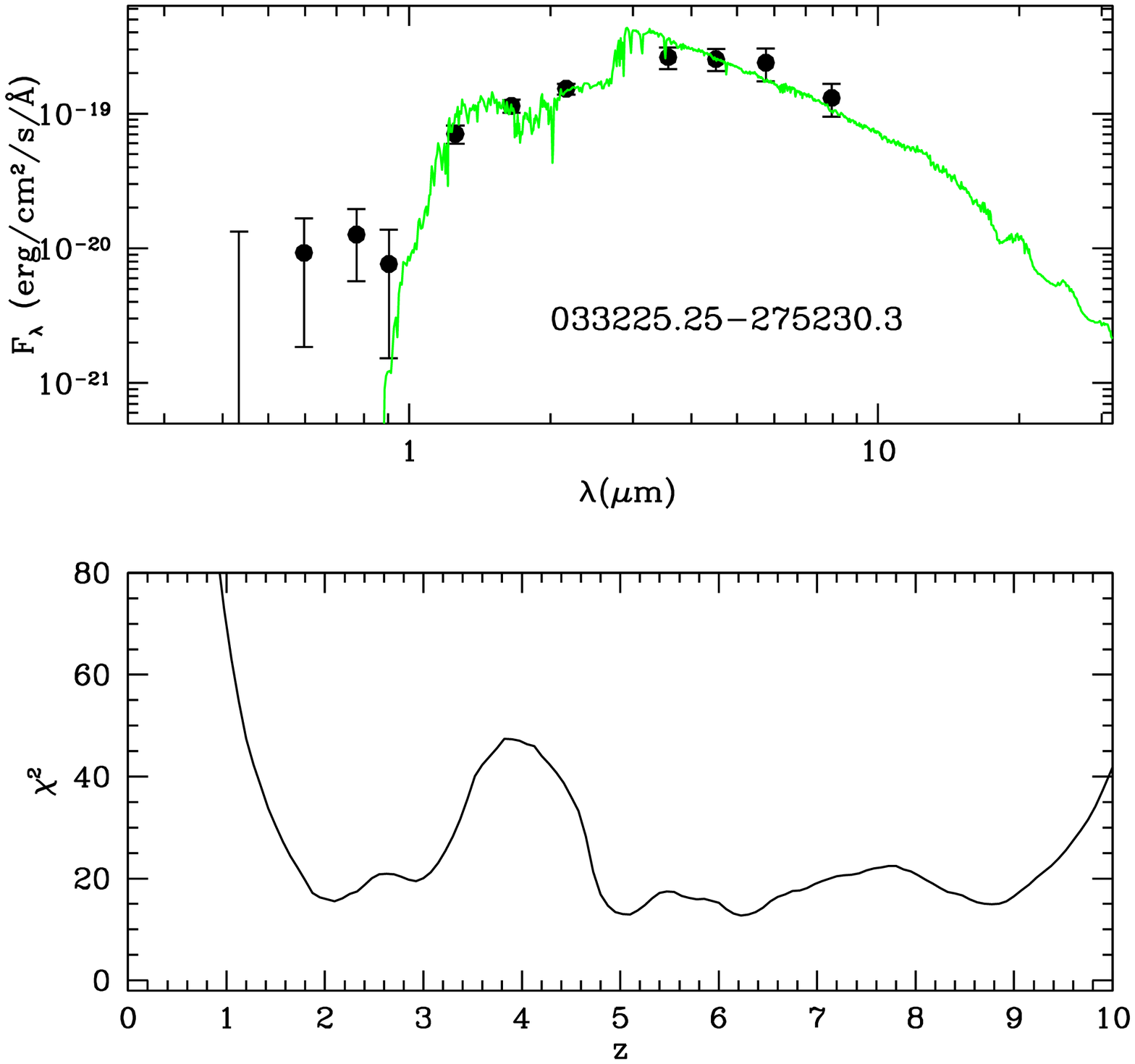}
\includegraphics{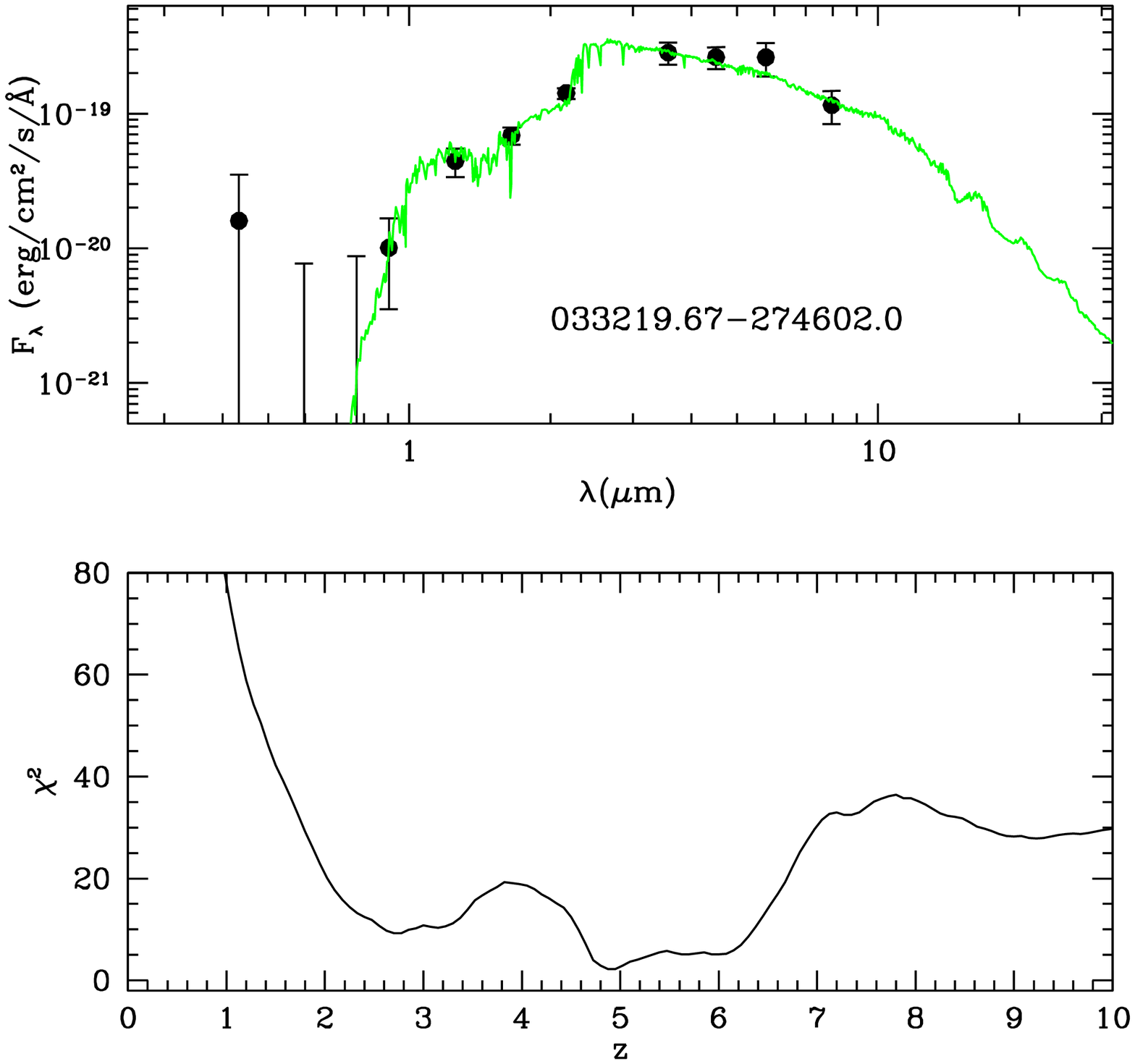}
\includegraphics{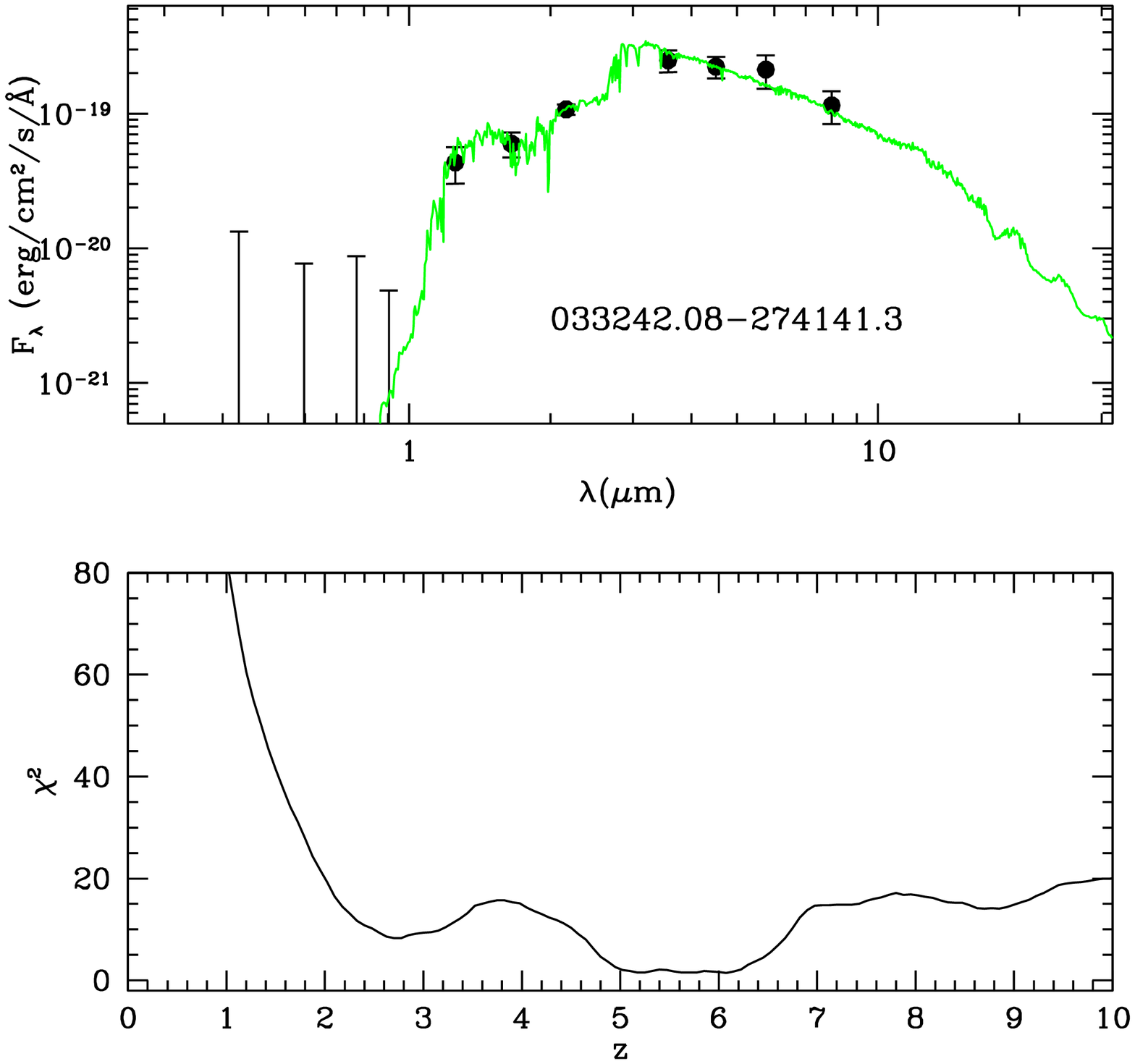}
\includegraphics{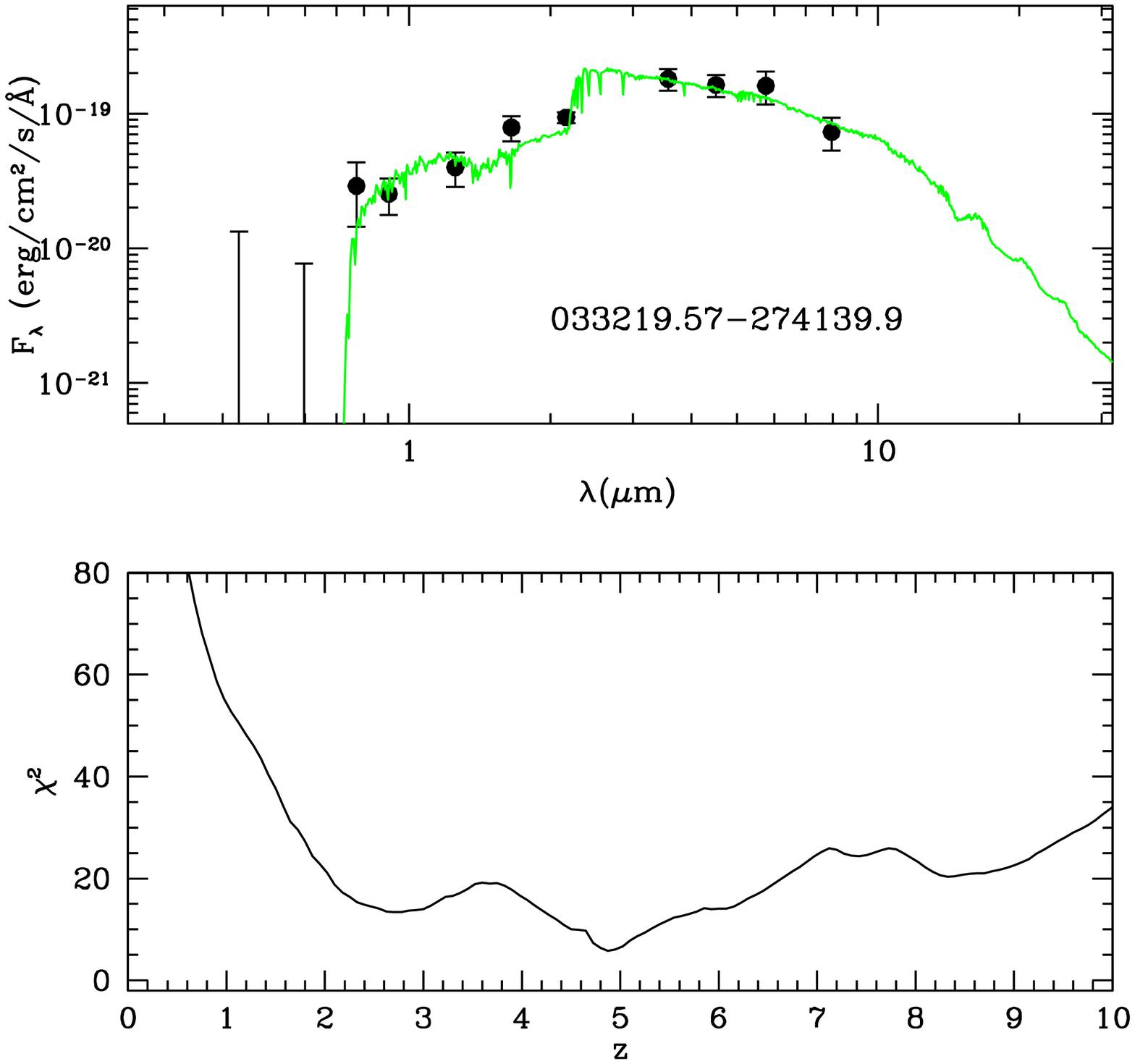}
\includegraphics{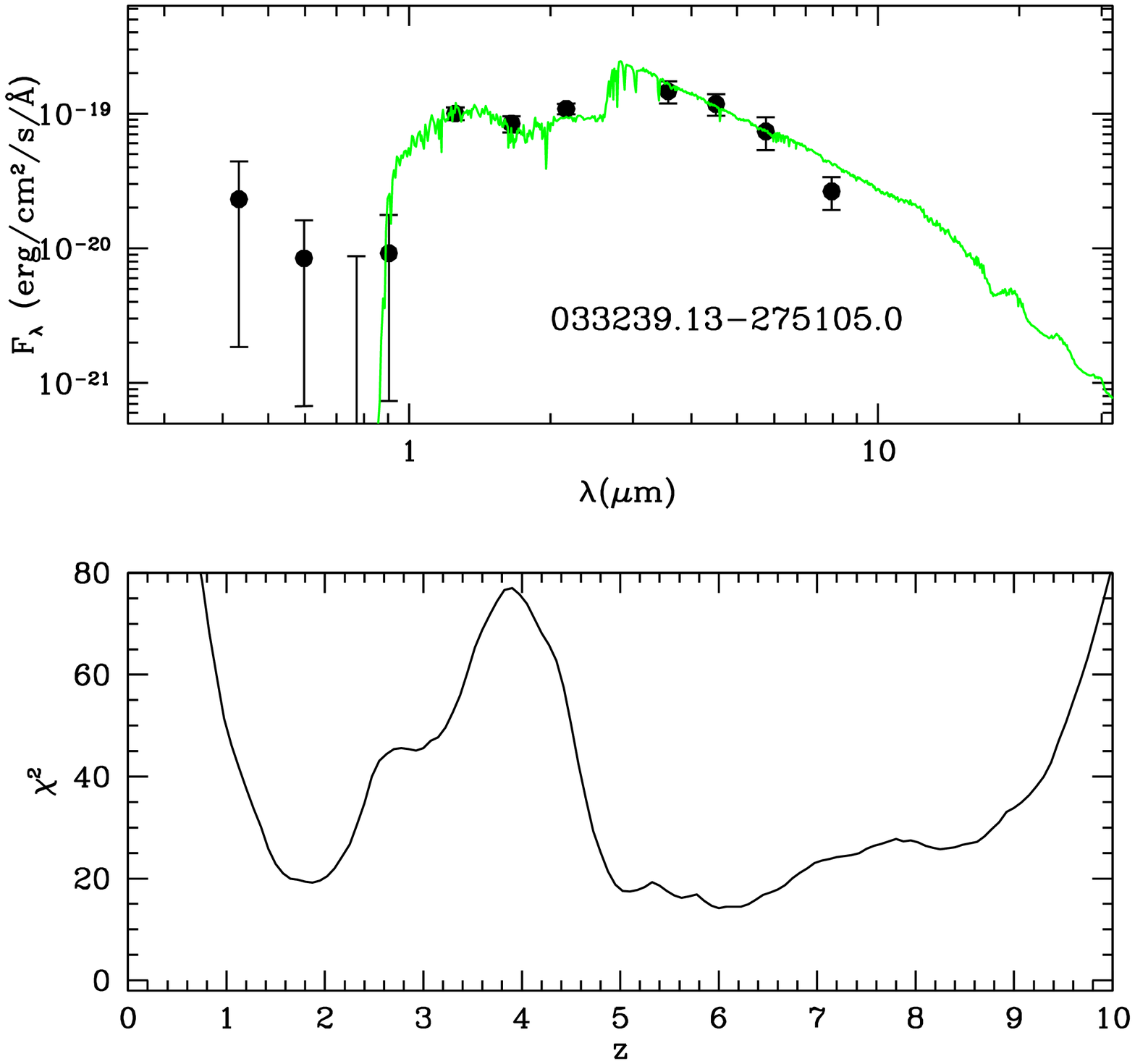}
\caption{\small Spectral fits and $\chi^2$ versus estimated redshift $z$
for the six galaxies in our GOODS-South sample which still have plausible
solutions at $z > 4$ after the sample refinement process described 
in Section 5. Best fitting model parameter values are given in Table 6.
For $A_V < 2$, the high-redshift solutions shown here are formally preferred.}
\end{figure*}

\section{Massive galaxies at high redshift?}

The model fits, and plots of $\chi^2$ versus redshift shown in Figure 4 were 
derived with dust extinction limited to $A_V < 2$. As detailed in Table 6, 
most of these `high-redshift' solutions are statistically acceptable, and 
yield plausible values for many of the model parameters ({\it e.g.} ages
less than the age of the universe at the epoch of interest). However, while
obviously not as surprising as the claimed discovery of HUDF-JD2, 
the existence of 6 galaxies at $z > 4$ with masses $M > 3 \times 10^{11} 
{\rm M_{\odot}}$ within a 125 square arcmin field is still unexpected.

Therefore, as a final step in the analysis we refitted these 6 galaxies 
with extinction now allowed to range up to $A_V \simeq 10$. The results 
of this process are illustrated in the contour plots shown in Figure 5. 
Acceptable, alternative low-redshift solutions are now found for all 6 
galaxies and, as detailed in Table 6, are now formally preferred for all 
except two galaxies (2507 and 3048), and even in these two cases the 
lower-redshift solution is formally acceptable. 

\begin{figure*}
\vspace*{15cm}
\includegraphics{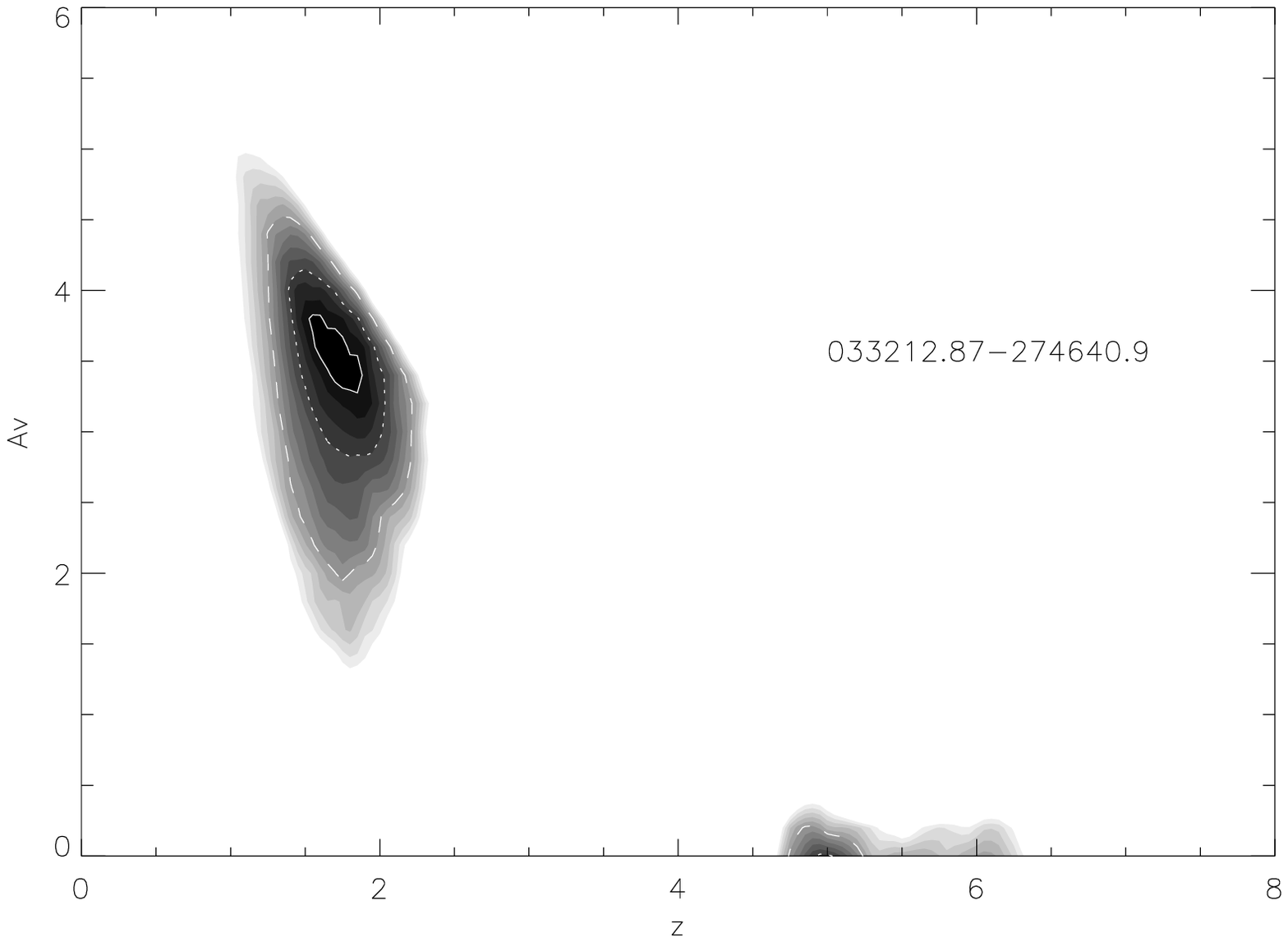}
\includegraphics{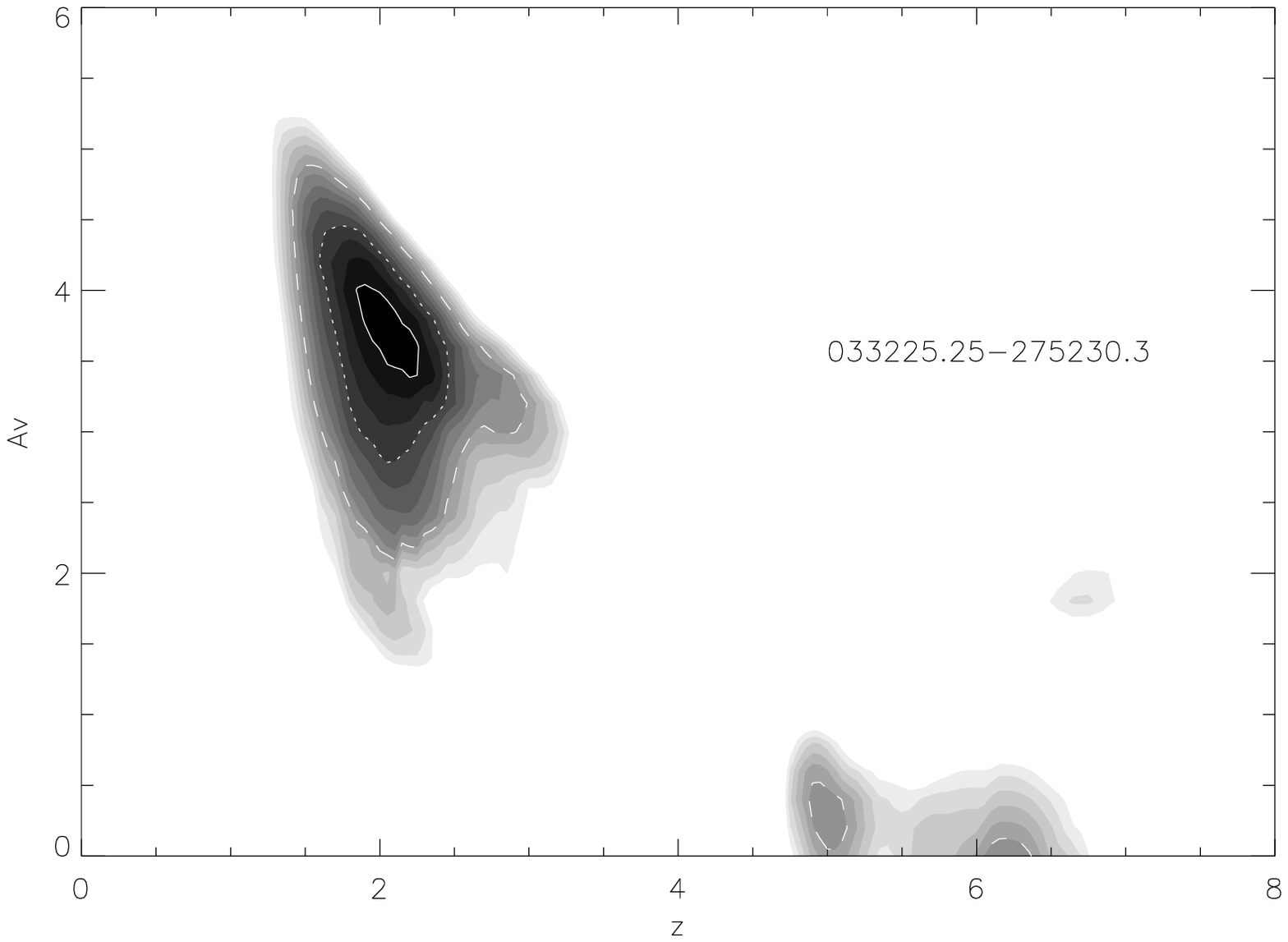}
\includegraphics{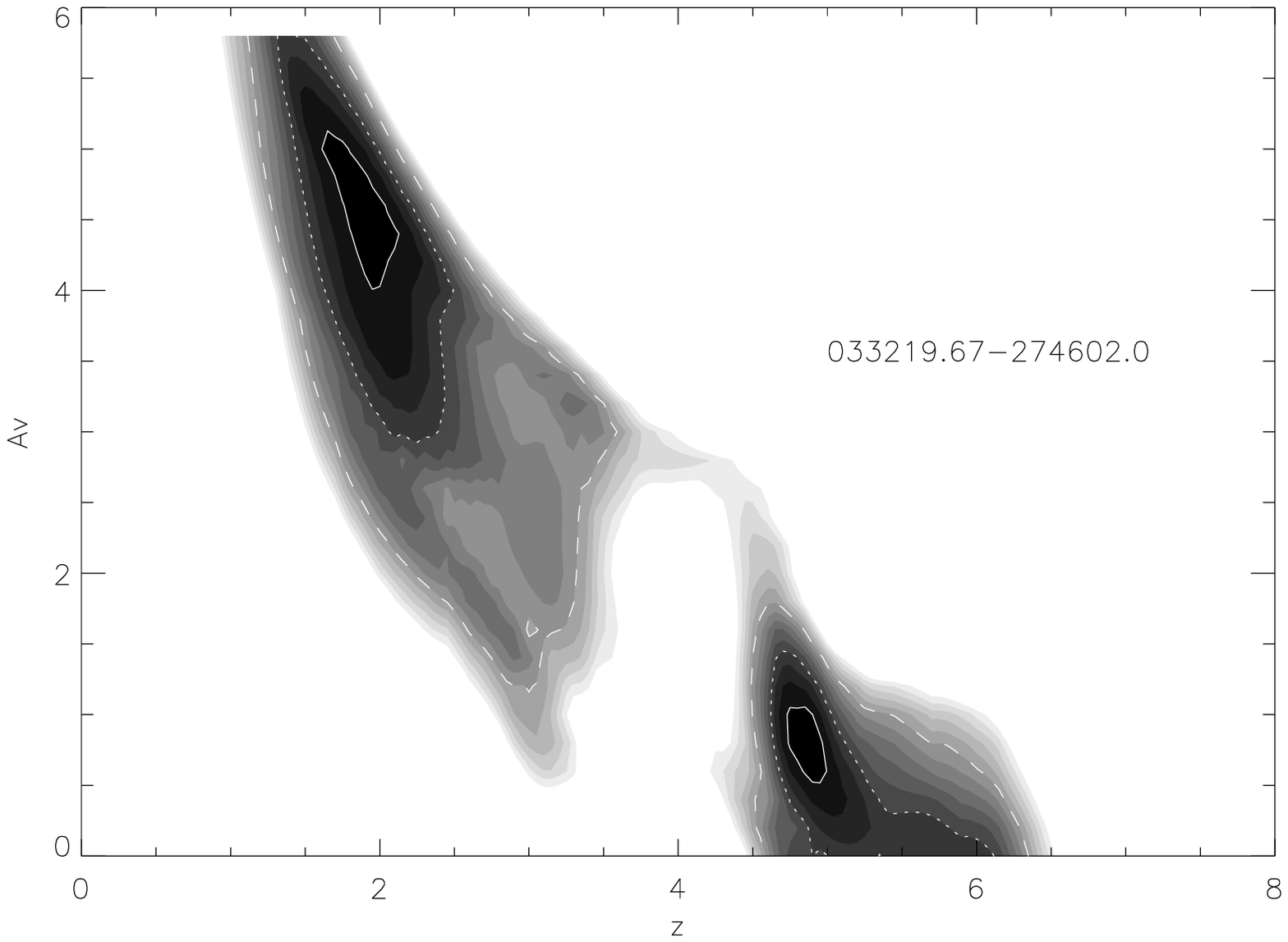}
\includegraphics{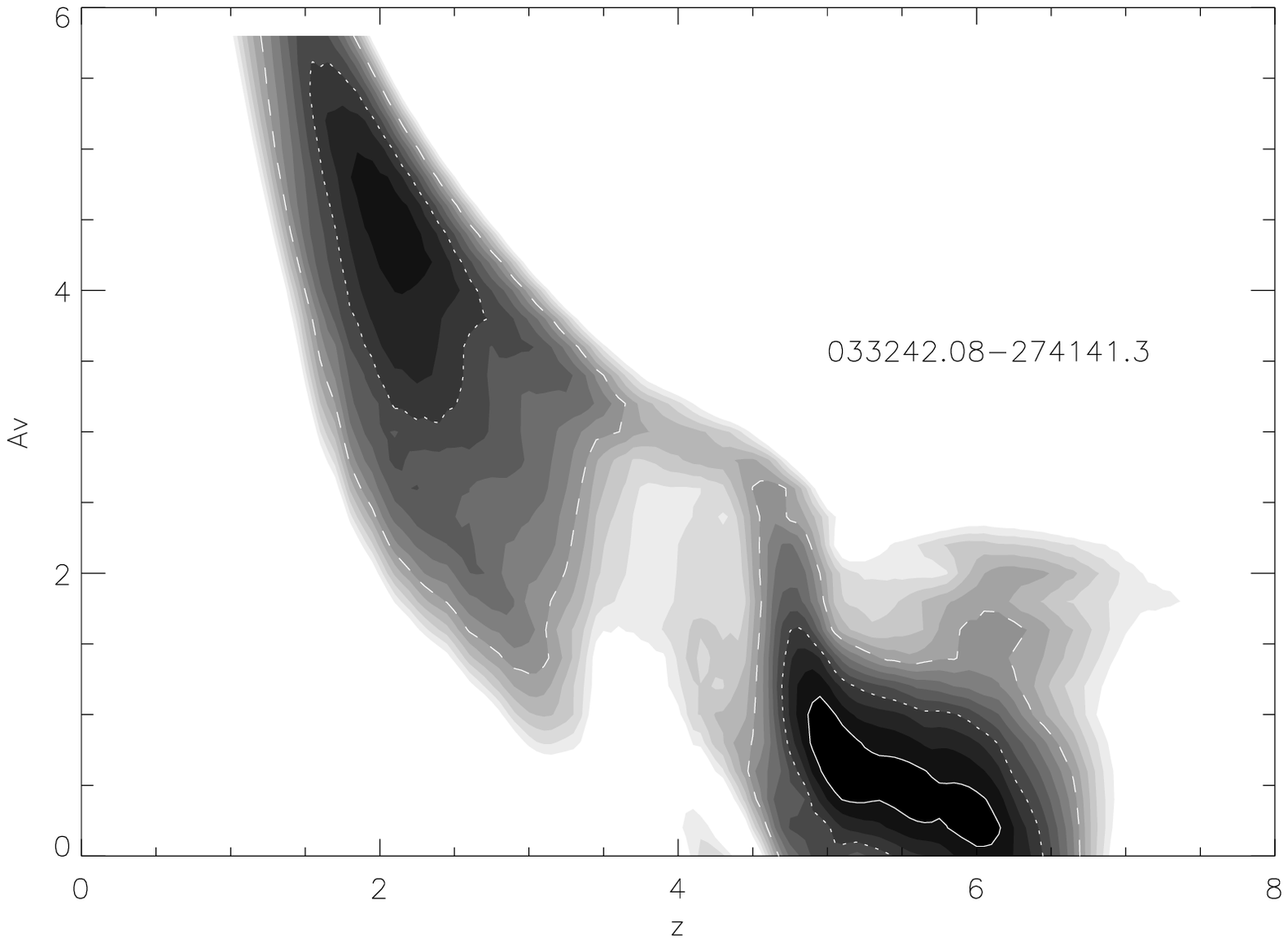}
\includegraphics{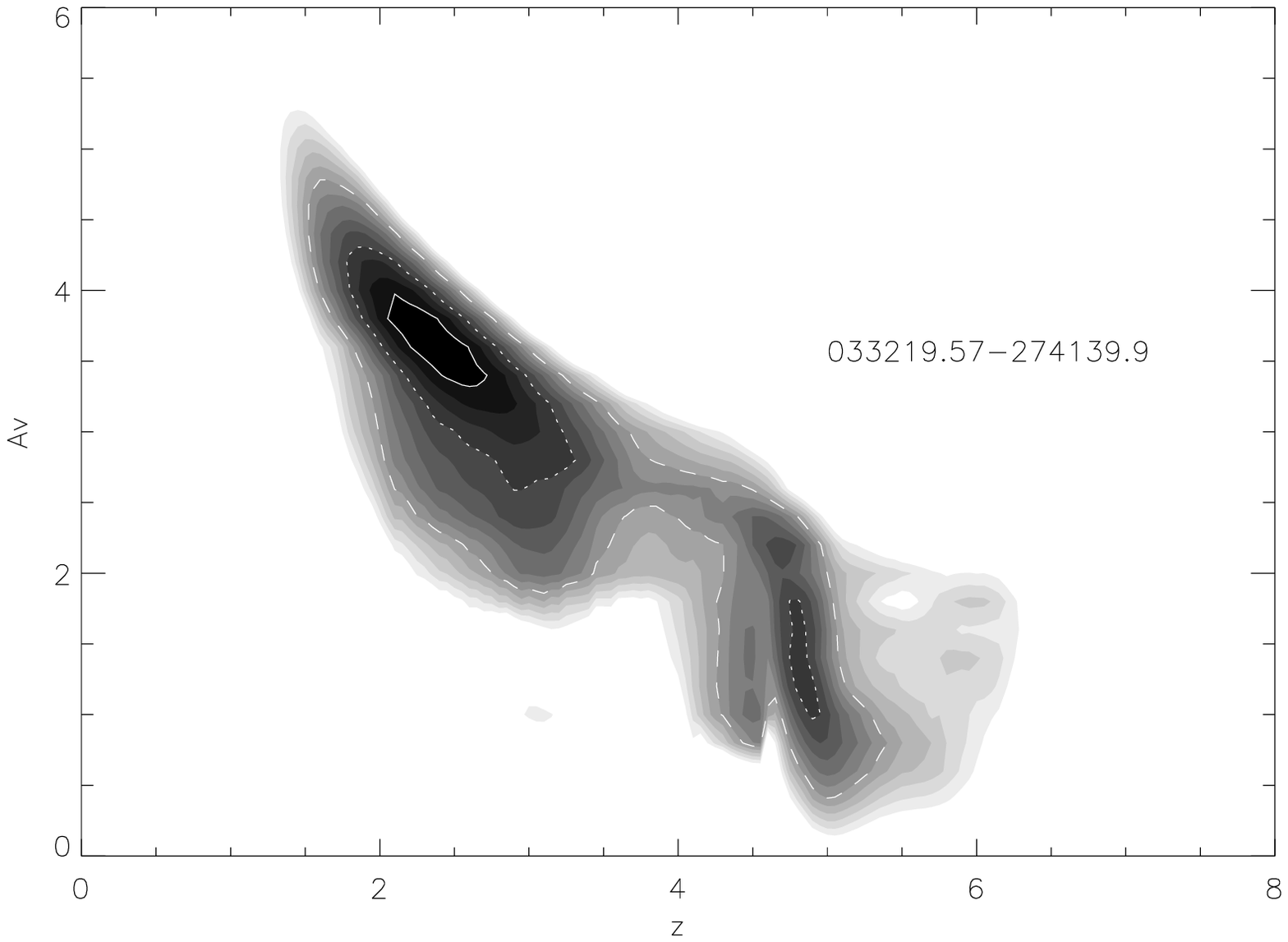}
\includegraphics{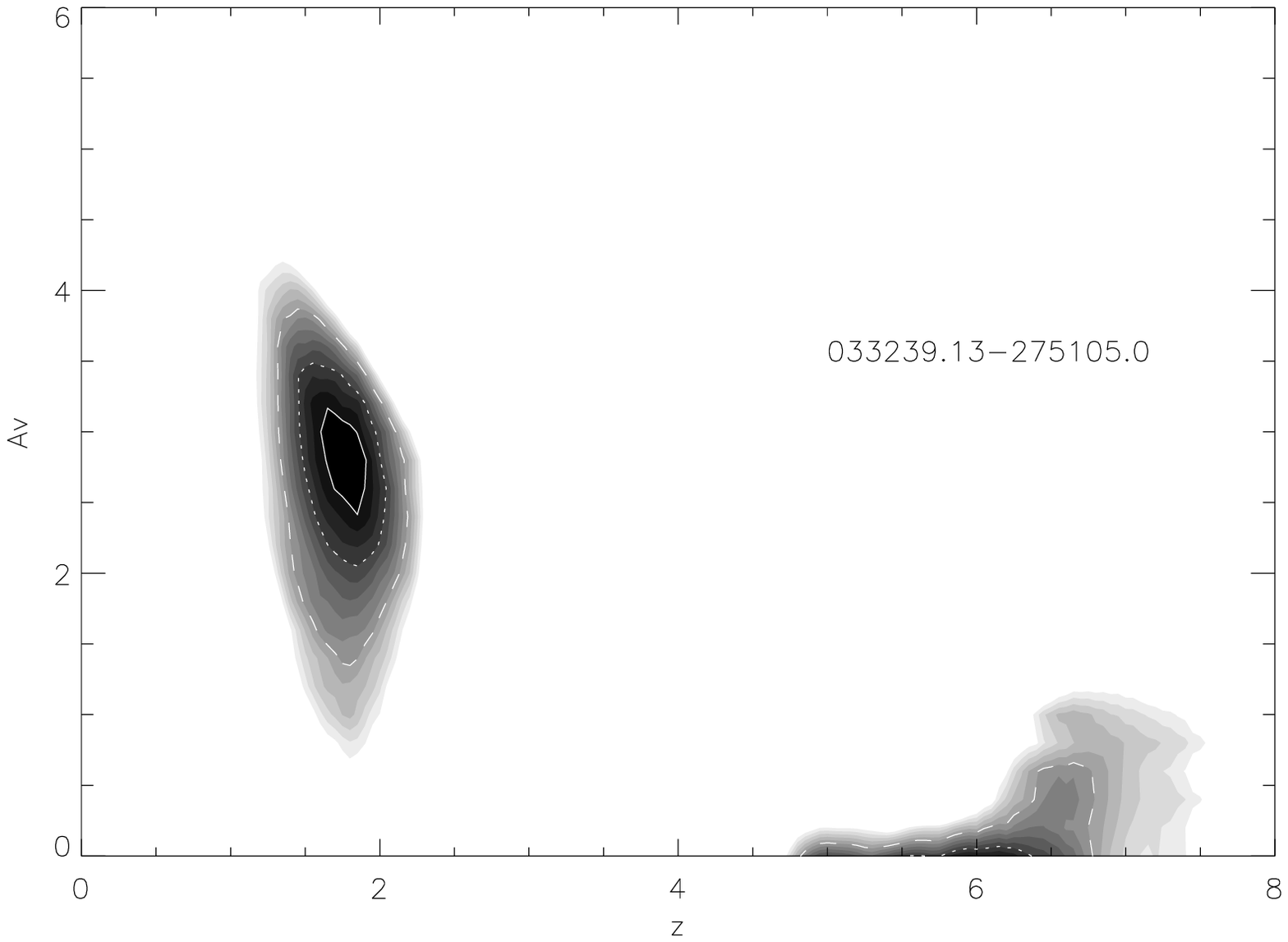}
\caption{\small Likelihood contours illustrating the location 
of acceptable fits on the $A_V$-$z$ plane for the 6 objects in our GOODS-South 
sample with putative
redshifts $z > 4$. Allowing for large values of extinction, it can be seen 
that, for all 6 objects, alternative solutions exist at $z \simeq 2$. 
Contours are shown at intervals of $\Delta \chi^2$ = 1, 4, and 9 above 
the minimum value of $\chi^2$. The best-fitting values of the model
parameters for both the high-redshift and low-redshift solutions 
are given in Table 6.}
\end{figure*}

It clearly remains possible that one or more of these 6 galaxies lies 
at extreme redshift, but a number of factors mitigate against this conclusion. 
First, given equally acceptable solutions at $z \simeq 2$ (with moderate mass 
but high $A_V$) and $z \simeq 5$ (with high mass and low $A_V$) the balance 
of other probabilities clearly favours the low redshift option. 
Second, during the completion of this work, 
the Spitzer MIPS 24$\mu m$ catalogue for the GOODS-South field was released
to the public (Dickinson et al., in preparation). 
This catalogue contains 24$\mu m$ detections 
for 5 out of 
the 6 galaxies listed in Table 6 (all except 3088). Clearly, such 
a high detection rate at mid-infrared wavelengths strongly supports 
the lower redshift dusty solutions for these objects (we note that HUDF-JD2 
was also detected at 24$\mu m$; Mobasher et al. 2005).
Third,
our own re-analysis of HUDF-JD2, and the discrepancy between the GOODS-MUSIC
redshift estimates and our own results for several of these putative 
high-redshift galaxies, serves to demonstrate just how sensitive any conclusion
in favour of $z > 4$ can be to the treatment of marginal and non-detections
in the optical wavebands. As demonstrated by Figure 1, this is 
a much less serious issue for young/blue high-redshift candidates for which 
an acceptable lower-redshift dust-obscured solution often does not exist.
However, the results presented here for redder objects serves to illustrate
just how hard it will be to unambiguously identify older (and hence potentially
more massive) objects at extreme redshifts on the basis of spectral 
fitting to even the most comprehensive photometric dataset.

\begin{table*}
\begin{center}
\caption{Best fitting model model parameter values for 
the final 6 candidate high-redshift objects. The values for the physical 
parameters, and the value of minimum $\chi^2$ is given 
for both the putative high-redshift (low $A_V$) and the moderate-redshift
(high $A_V$) solution in each case. The locations of these solutions on the 
$A_V - z$ plane are illustrated in Figure 5.}
\begin{tabular}{lccccrccc}
\hline
Source & RA (J2000) & Dec (J2000) & ${\rm z_{est}\,\,(1\sigma\, range)}$ 
& Model type & $\chi^2$ &Age/Gyr & ${\rm A_V}$ & Mass/$10^{11}\,{\rm M_{\odot}}$  \\
\hline
\\
1865 & 03 32 12.87 & $-$27 46 40.9& 5.02\, $(4.87-5.17)$& Burst & 8.82 & 0.56 &0.0 &\phantom{1} 6.0\\
     &&& 1.75\, $(1.55-1.95)$&Burst & 5.20 & 0.13 & 3.6 & \phantom{1}0.9\\  
2336 & 03 32 25.25 & $-$27 52 30.3& 6.22 $(6.07-6.45)$& Burst & 12.73&   0.56 &0.0 &16.6\\   
     &&& 2.20\, $(1.85-2.35)$&Burst & 4.80 & 0.13 & 3.6 & \phantom{1}2.0\\

2507 & 03 32 19.67 & $-$27 46 02.0&4.87 $(4.72-5.10)$& Burst & 2.20 &   0.57 &0.8 &11.2\\         
     &&& 1.85\, $(1.55-2.30)$& $\tau = 15$\, Gyr & 2.43 & 0.20 & 4.8 & \phantom{1}0.8\\
2958 & 03 32 42.08 & $-$27 41 41.3&6.07 $(4.95-6.30)$& Burst & 1.49 &   0.63 &0.2 &15.8\\         
     &&& 2.05\, $(1.75-2.40)$ & Burst & 3.03 & 0.18 & 4.6 & \phantom{1}1.9\\
3048 & 03 32 19.57 & $-$27 41 39.9&4.87 $(4.72-5.10)$& Burst & 5.83 &   0.31 &1.2 & \phantom{1}5.9\\    
     &&&2.45\, $(2.10-2.80)$  & $\tau = 0.3$\, Gyr & 2.39 & 0.07 & 3.6 & \phantom{1}0.4  \\
3088 & 03 32 39.13 & $-$27 51 05.0&6.00 $(5.85-6.37)$& Burst & 14.14&   0.40 &0.0 & \phantom{1}2.8\\    
     &&& 1.90\, $(1.65-2.00)$ & Burst & 12.46 & 0.20 & 2.8 & \phantom{1}0.6\\\hline
\end{tabular}
\end{center}
\end{table*}

\section{Discussion}

To illustrate the implications of this failed search for high-mass galaxies at
$z > 4$ we show, in Figure 6, a plot of the comoving number density
of galaxies with mass $M > 3 \times 10^{11} {\rm M_{\odot}}$ within the 
GOODS-South field, as a function of redshift. This limit simply corresponds 
to the lowest mass found for the high-redshift solutions 
to the 6 galaxies listed in Table 6. 

The data points for the 
redshift bins $0 < z < 1$, $1 < z < 2$, $2 < z < 3$, and $3 < z < 4$ are 
derived from the analysis of Cirasuolo et al. (2006) (updated from Caputi
et al. 2006), while the reference point at $z = 0$ has been derived by 
the appropriate integration of the galaxy mass function provided by Cole et al.
(2001) (assuming a Salpeter IMF). 
The result of the unsuccessful search at $z > 4$ described here is 
illustrated by the upper limit plotted at $z = 5$, which represents 
the comoving number density in the redshift bin $4 < z < 6$ if {\it one} 
of the 6 final high-redshift candidates listed in Table 6 really does lie
within this high-redshift bin.
Also plotted in this figure are the comoving number density of virialized 
dark-matter halos above 3 different mass thresholds, as derived from modified
Press-Schechter theory (Percival, priv. comm.). 

Several features of this diagram are worthy of comment.
First, at no redshift does the number density of high-mass galaxies 
present a fundamental problem for $\Lambda$CDM; the number density of 
potential dark-matter halos (10-20 times more massive than the stellar masses 
of the galaxies) clearly
exceeds the inferred number density of the massive galaxies. 

Nevertheless, out to redshift $z \simeq 3$, the number density of 
these most massive galaxies changes only slowly (compared to the dark matter
curves) and hence, for this class of galaxies, 
the inferred dark-matter:stellar mass ratio appears to evolve from $\simeq
80$ at $z \simeq 0$ to $\simeq 20$ at $z \simeq 3.5$. Such inferred values
do not appear unreasonable (see, for example, Mandelbaum et al. 2006), 
and this apparent
evolution of the dark-matter:stellar mass
ratio can be viewed as yet another manifestation of `downsizing'
or apparently `anti-hierarchical' galaxy formation (e.g. Heavens et al. 2004).

However, the upper limit derived here indicates that 
this `anti-heirarchical' behaviour does not persist beyond $z \simeq 4$, and 
that, at higher redshift, the number density of massive galaxies drops 
off as rapidly (or possibly more rapidly) than the number density of 
potential host dark-matter halos.

An improved measurement of the comoving number density of these rare high-mass
galaxies at high-redshift clearly requires a substantially larger survey than 
the 125 square arcmin covered by the GOODS-South multi-wavelength imaging.
The first such survey of the necessary depth and area is now underway. This 
is the UKIDSS Ultra Deep Survey (UDS), 
which covers 0.8 square degrees, and is 
designed to ultimately 
reach a $K$-band 5-$\sigma$ detection limit of $K = 25$ (AB). 
This survey will therefore cover 25 times the area of GOODS-South, to
a magnitude limit substantially deeper than the $K_S$-limit of the GOODS-South 
survey analysed here. A failure to find any galaxies 
with $M > 3 \times 10^{11} {\rm M_{\odot}}$ and $z > 4$ within the UKIDSS
UDS would move the upper limit shown at $z \simeq 5$ in Figure 6 down by 
over an order of magnitude. 
A first analysis, based on the UKIDSS early data release,
has provided evidence for a number of moderately 
massive galaxies at $ z > 5$, but as yet has not revealed 
any as massive as $M > 3 \times 10^{11} {\rm M_{\odot}}$ (McLure et al. 2006).

\begin{figure}
\vspace*{7.5cm}
\includegraphics{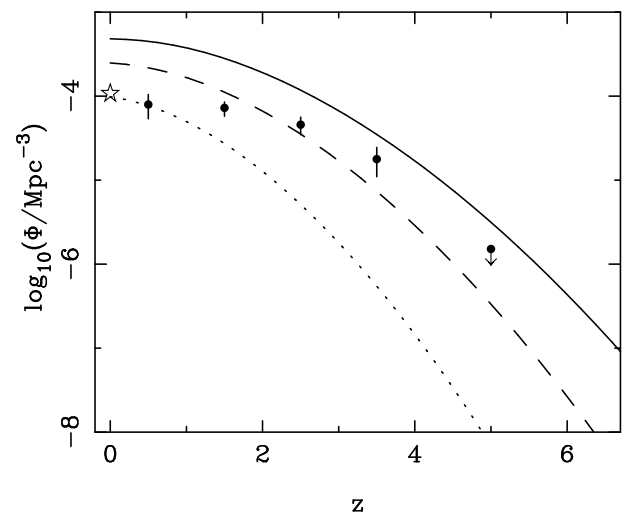}
\caption{\small A comparison of our best estimate 
of the evolution of the comoving number density
of galaxies with stellar mass $M > 3 \times 10^{11} {\rm M_{\odot}}$ 
with the predicted evolution of the 
comoving number density of virialized 
dark-matter halos above 3 different mass thresholds (as derived from modified
Press-Schechter theory; Percival, priv. comm.). 
The data points for the 
redshift bins $0 < z < 1$, $1 < z < 2$, $2 < z < 3$, and $3 < z < 4$ are 
derived from the analysis of Cirasuolo et al. (2006) (updated from Caputi
et al. 2006), while the reference point at $z = 0$ has been derived by 
the appropriate integration of the galaxy mass function provided by Cole et al.
(2001) (assuming a Salpeter IMF). 
The result of the unsuccessful search at $z > 4$ described here is 
illustrated by the upper limit plotted at $z = 5$, which represents 
the comoving number density in the redshift bin $4 < z < 6$ if {\it one} 
of the 6 final high-redshift candidates listed in Table 6 really does lie
within this high-redshift bin.
The curves show the comoving number density of virialized 
dark-matter halos with $M > 5 \times 10^{12} {\rm M_{\odot}}$ (solid line),
$M > 1 \times 10^{13} {\rm M_{\odot}}$ (dashed line),
and $M > 2.5 \times 10^{13} {\rm M_{\odot}}$ (dotted line).
The last (most massive) of these curves was deliberately chosen to coincide 
with the data-point at $z = 0$, and implies a dark-matter:stellar mass 
ratio of $\simeq 80$ for these most massive galaxies in the present-day 
universe.} 
\end{figure}

\section*{ACKNOWLEDGEMENTS}
This work was based in part on 
observations with the NASA/ESA {\it Hubble Space Telescope},
obtained at the Space Telescope Science Institute, which is operated by the 
Association of Universities for Research in Astronomy, Inc. under NASA 
contract No. NAS5-26555. Michele Cirasuolo acknowledges the support of PPARC,
on rolling grant no. PPA/G/O/2001/00482.
Ross McLure acknowledges the support of the Royal Society, through the award
of a Royal Society URF.

\end{document}